\documentclass[natbib]{svjour3}                     
\smartqed  
\usepackage{times, graphicx}
 \usepackage{aps-bibstyle}  
\usepackage{latexsym}
 \journalname{Space Science Reviews}
\newcommand{\bfxi}{\mbox{\boldmath$\xi$}}

\begin{document}

\title{Alfv\'en Waves in the Solar Atmosphere}
\subtitle{From Theory to Observations}


\author{M. Mathioudakis, D. B. Jess \& R. Erd\'elyi}


\institute{M. Mathioudakis \& D. B. Jess \at
              Astrophysics Research Center\\
              School of Mathematics and Physics\\ 
              Queen's University Belfast\\ 
              Belfast BT7 1NN, Northern Ireland, UK \\
              Tel.: +44 28 9097 3573\\
              Fax: +44 28 9097 3997\\
              \email{m.mathioudakis@qub.ac.uk, d.jess@qub.ac.uk}           
           \and
           R. Erd\'elyi  \at
            Solar Physics \& Space Plasma Research Center (SP$^2$RC)\\
          School of Mathematics and Statistics\\
           University of Sheffield\\
           Sheffield S3 7RH, UK\\
           \email{robertus@sheffield.ac.uk}
           }

\date{Received: date / Accepted: date}

\maketitle

\begin{abstract}
Alfv{\'e}n waves are considered to be viable transporters of the non-thermal energy required to 
heat the Sun's quiescent atmosphere. An abundance of recent observations, 
from state-of-the-art facilities, have reported the existence of Alfv\'en 
waves in a range of chromospheric and coronal structures. Here, we 
review the progress made in disentangling the characteristics of transverse 
kink and torsional linear magnetohydrodynamic (MHD) waves. We outline the 
simple, yet powerful theory describing their basic properties in 
(non-)uniform magnetic structures, which closely resemble the building blocks 
of the real solar atmosphere.

\end{abstract}

\section{Introduction}
\label{intro}
The solar coronal plasma is an ionized gas which is mainly confined by the ubiquitous 
presence of magnetic flux tubes and open magnetic field lines. These structures 
demonstrate sizes over very large scales, right down to the current observational limit, 
and are maintained at temperatures 
of several million Kelvin. The heating processes that generate, and sustain the hot 
corona have so far defied quantitative understanding, despite efforts spanning more 
than half a century \citep{kup_etal81,gom90,zir93,ofman05,kli06,tar-erd09}. 
Efforts to establish the causes of solar atmospheric heating 
have produced a number of theories, of which two classes are most promising. 
The first holds for current dissipation, 
following reconnection events occurring throughout the atmosphere in the 
form of micro-flare and nano-flare activity \citep{RefA,pri-sch99,fuj_etal11}. The second class predicts 
that the heating is driven by magneto-hydrodynamic (MHD) waves  \citep{alf47,osterbrock61,ion78,hol91}, most likely 
propagating from the lower atmosphere \citep{RefB}. 
Turbulence can transfer energy from large scales to small scales.  
There is growing  evidence  that MHD turbulence and small-scale reconnections are explicitly linked, with
current sheets acting as the dissipative structures. Magnetic reconnection is also observed in simulations 
of MHD turbulence \citep{mat11,rappazzo08}. Two and three dimensional simulations show that magnetic 
structures tend to self-organize in space and time, creating a number of current sheets at their boundaries. 
The intermittency of microflares and nanoflares, combined with the small spatial scales over which the 
energy release occurs, allows us to refer to them collectively as turbulence \citep{einaudi96,georg98}.

Kinetic processes and models that are based on wave particle interactions can not explain yet solar phenomena in a global scale, but are clearly applicable to small spatial scales. Kinetic processes and instabilities are highly relevant in models of plasma heating and solar wind acceleration. Hybrid models consider the acceleration of heavy ions by a spectrum of waves in the solar wind as well the influence of the ions on the wave structure and can reproduce the anisotropic velocity distributions of heavy ions 
\citep{scu93a,scu93b,mat11,ofman07}.
 
A strongly related, and also unsolved question, involves the generation and acceleration 
of the solar wind, which often violently interacts with the magnetosphere and Earth's 
upper atmosphere. Again, there are strong arguments that MHD waves play a key role 
in contributing to these important questions of space physics \citep{belcher71,alazraki71,gek99,tsu99}. The 
compelling theory of Alfv\'en waves, which is the basis and starting point of most 
of the above seminal work, was developed by 
Hannes Alfv\'en \citep{alf42}, for which he received a Nobel Prize in 1970.

Fast and slow mode magneto-acoustic waves cause compression and rarefaction 
of the plasma as they propagate, with the magnetic and gas pressures acting as 
the main restoring forces, although kink modes are highly incompressible and the dominating 
restoring forces are magnetic tension-driven. Alfv\'en waves 
are incompressible, with magnetic tension the only restoring force for linear perturbations. 
If the perturbations are non-linear in nature, magnetic tension is the dominant restoring force.
Alfv\'en waves will cause primarily Doppler shifts, or specific magnetic field 
perturbations detectable with spectro-polarimetry in selected magnetically sensitive 
lines. Magneto-acoustic waves can cause both intensity variations and Doppler shifts.
If the oscillating structure is spatially unresolved, the above effects will result in a 
broader line profile.

A study of the dynamic properties of the interplanetary medium, carried out by the 
Mariner 5 spacecraft in 1967, has shown large-amplitude Alfv\'en waves in 
micro-scale structures \citep{colem67,unti68}. The largest amplitudes were identified in high velocity streams, 
but were also detected in low velocity regions. It is suggested that these are the 
remnants of waves generated near the solar surface, and have managed to propagate 
into the heliosphere without significant damping. The non-detection of appreciable 
magneto-acoustic wave power would suggest that these waves may have dissipated 
their energy in the chromosphere and corona \citep{RefBel}. The Mariner observations, combined with the polar observations carried out by Ulysses, allowed a comprehensive view of turbulence in the heliosphere. Measurements of the solar wind parameters in the ecliptic, have shown that the  Alfv\'enic turbulence observed in high velocity streams, evolves towards a Kolmogorov-type spectrum with velocity shear helping to drive the process. The polar observations reveal a similar evolution but at a lower pace \citep{bruno05}. These observations have prompted the development  of wave-driven models to explain wind acceleration and heating \citep{velli03} (see \cite{ofman10} for a recent review).

The motivation for the search for Alfv\'en waves in the corona can be traced back to 
\cite{alf47}. Observations of Alfv\'en waves in the magnetosphere \citep[e.g.][]{RefKeiling} 
have prompted the search for the source  of these waves in the lower solar atmosphere. 
With the advent of state-of-the-art instruments on-board satellite missions and 
ground-based solar telescopes, Alfv\'en  and transversal kink wave modes have started 
to be studied in the solar photosphere, chromosphere and corona at an unprecedented 
level of detail \citep {asch99,nakariakov99,ofman08,RefAx}. The primary objective of this paper is to give an unbiased, but unavoidably 
personal account of the recent observational reports of Alfv\'en waves in the solar atmosphere. 
This work will be 
underpinned by the linear magneto-hydrodynamic (MHD) theory that describes these waves in 
(non-)uniform magnetic structures (e.g. slabs and cylinders), which are considered to be simple, 
yet accurate descriptions of the actual magnetic building blocks of the solar atmosphere. We point out that 
the accuracy of slabs, cylinders, and pure linear modes in the description of realistic coronal 
active region loops is often doubted. Improved 3D MHD computational studies of coronal loops, 
in more realistic bi-polar active region geometries, has been recently performed by 
\citet{mclaughlin08}, \citet{ofman09a}, \citet{selwa10}, \citet{schmidt11}, and \citet{selwa11a,selwa11b}.

\section{The Transition Region and Corona}
\label{sec:1}

\subsection{Non-thermal Line Broadening} 

The global torsional oscillation of a coronal loop, with an axis that is not parallel to the 
line of sight, can produce periodic variations in the observed spectral line widths 
\citep{RefE,RefF,RefAw}. The first observational efforts to detect Alfv\'en waves in transition 
region (TR) and coronal structures dates back to the 1970s, with observations coming from 
the SKYLAB and SMM missions. Analysis of UV spectra, obtained with the Naval 
Research Laboratory spectrograph, revealed non-thermal velocities in the range 
of 10--30 km{\,}s$^{-1}$ \citep{RefC, RefD}. The main conclusion from these 
studies is that, if the excess line broadening is to be attributed to Alfv\'en waves, 
these waves cannot account for the heating. 
\cite{RefE} suggested that unambiguous evidence for Alfv\'en waves in the TR 
can be provided by tracking coronal loops as they move from disk center to the limb. 
The non-thermal widths of emission lines from these loops would be expected to 
increase as a function of distance from disk centre. These line widths should 
become maximum at the limb, where the magnetic field lines are predominantly 
perpendicular to the observer's line of sight. Indeed, SUMER observations have 
shown a centre to limb increase in the non-thermal line widths of upper 
chromospheric and TR lines, while the excess broadening of coronal lines is only 
marginal (see e.g. \citealt{RefF,Ban_etal07} and references therein). However, 
chromospheric and transition region lines are 
known to suffer from opacity effects. Thus, the center-to-limb variation of the 
line widths can also be attributed to increased opacity broadening towards the limb.  

The line broadening could also be due to the effects of bulk outflow motions that cause 
differential Doppler shifts along the observer's line-of-sight. The height profile of the temperature 
distribution, bulk outflow motions, and the geometry of superradial expansion will 
enhance the Doppler broadened line profile.  These effects are exacerbated by an 
anisotropic temperature distribution, particularly when 
T$_{\perp}$ $>>$ T$_{\parallel}$ \citep{akinari07, ofman10a}.
\cite{raouafi06} also find that at  distances greater that 1R$_{\odot}$ from the 
solar surface, the widths of some transition region and coronal lines depend on 
the adopted electron density stratification. This became apparent in polar coronal 
hole observations of O~{\sc{vi}} and Mg~{\sc{x}} lines. This may indicate that the 
need for anisotropic temperature/velocity distributions may not be as important 
as previously thought.

\begin{figure}
\includegraphics[width=0.99\textwidth]{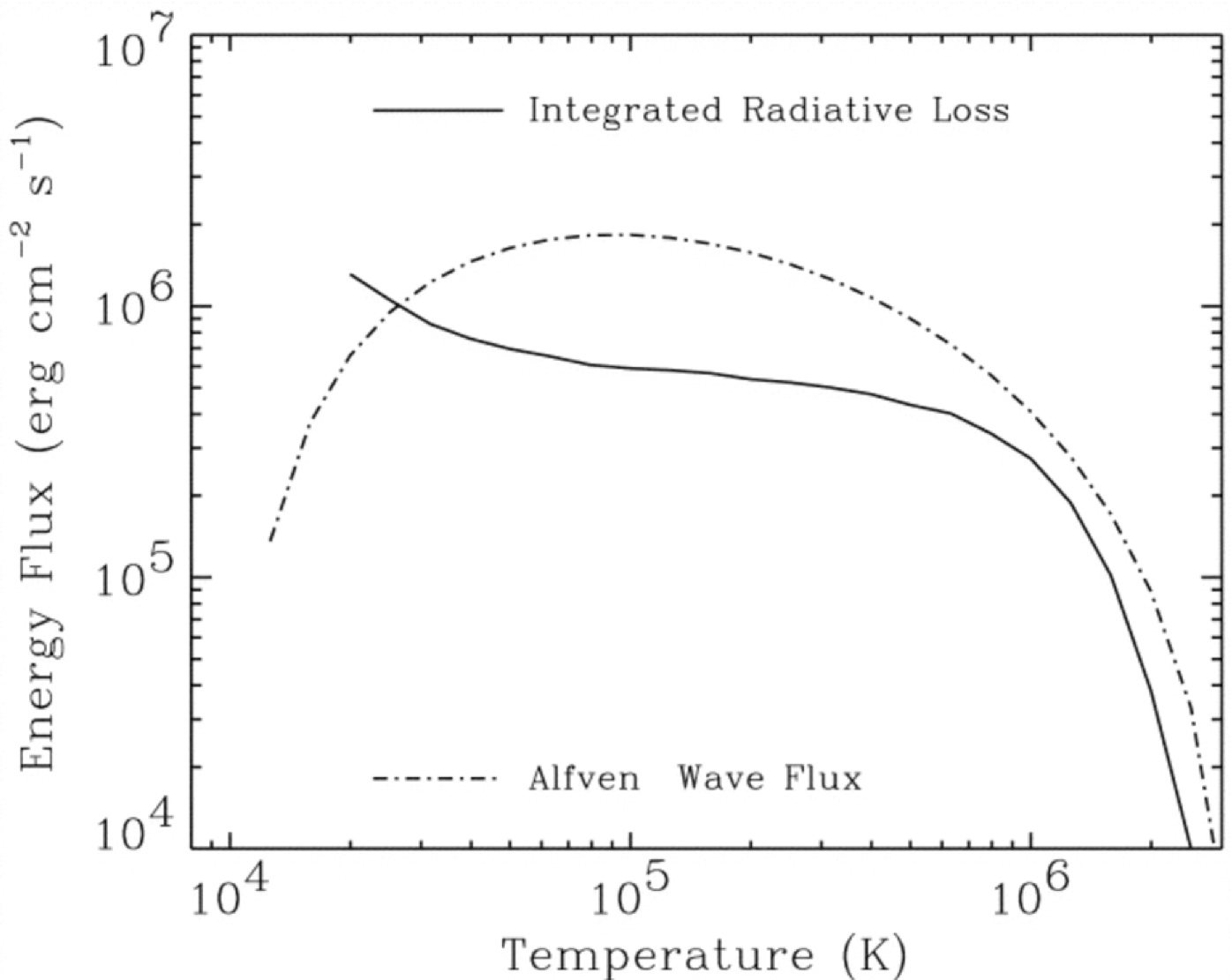}
\caption{The Alfv\'en wave energy flux compared with the radiative losses from the 
TR and corona \citep[reproduced from][]{RefG}. }
\label{fig:1}       
\end{figure}

In a follow up investigation, \cite{RefG} argue that the small center-to-limb 
variation can be due to network cell structures where the non-thermal velocities 
may be considered isotropic. The authors concluded that if the small increase 
in the non-thermal velocities of EUV lines is due to Alfv\'en waves, the limited 
temporal resolution of the observations would imply that the wave periods will 
be shorter than a few tens of seconds. They used the non-thermal velocities 
to derive a wave flux, which was then compared with the radiative losses in the 
10$^4$ -- 2$\times$10$^6$~K temperature range. Their comparison reveals that 
there is sufficient Alfv\'en wave flux to balance the radiative losses for temperatures 
in excess of  3$\times$10$^4$~K. For temperatures lower than 3$\times$10$^4$~K, 
the wave energy is significantly smaller than the radiative losses. However, this is 
inconsistent with Alfv\'en wave flux which is expected to remain constant 
(or decrease) as the temperature increases \citep{RefG}. This rapid increase 
in energy as a function of temperature is a direct consequence of the small 
increase in the non-thermal velocities found in this temperature range (Fig.~\ref{fig:1}).   

\subsection{The CoMP View of the Corona}
\label{sec:2}

Direct measurements of the coronal magnetic field have been uncertain, and 
notoriously difficult. \cite{RefI} have used the near infrared magnetically sensitive 
Fe XIII 10747{\,}\AA\ line to measure the Stokes V profiles above active regions, 
where they determined longitudinal magnetic field strengths in the range of 10 -- 33~G.   

The state-of-the-art Coronal Multi-channel Polarimeter \citep[CoMP;][]{Dar03} 
has been employed behind a coronagraph at the Hill-top facility at the National 
Solar Observatory in New Mexico, to provide the states of polarization in the 
10747{\,}\AA\ line \citep{RefJ}. The observing setup allows measurements of spectral 
line intensities, line-of-sight velocities, and line widths in the corona between 1.05 
and 1.35 solar radii. The observations show clear evidence for line-of-sight velocity 
variations with a period of approximately 5 min (3.5 mHz). Simultaneous measurements 
of linear polarization have revealed the plane-of-the-sky azimuthal component of the 
coronal magnetic field. \cite{RefJ} suggested that these oscillations are the signatures 
of Alfv\'en waves in the corona for the following reasons: (a) The waves have phase 
speeds of about 2000~km{\,}s$^{-1}$, which is considerably higher than the coronal 
sound speed, (b) The waves propagate along density structures that are thought to follow the magnetic field lines, and (c) 
There is no evidence for appreciable intensity oscillations which indicates that the 
detected wave modes are incompressible. The lack of intensity oscillations may be due 
to the low spatial resolution that would average out any small scale variations (see later).

The Fourier spectrum of the Doppler velocities associated with these waves points 
to a photospheric driver (i.e. p-modes). However, since their frequency is below the 
chromospheric cut-off, the only way that they can propagate upwards into the corona is 
through inclined magnetic structures \citep{RefK,erd06a,erd06b}. They would also 
need to be converted from pressure-driven modes into non-compressible transversal 
Alfv\'en modes.

The interpretation of the CoMP results, in terms of Alfv\'en waves, has been criticised by e.g. \cite{RefL}, 
who argue that these observations show the collective motion of the plasma from its 
equilibrium position, thus causing a periodic variation in the line-of-sight velocity. The authors suggest 
the best interpretation is the fast kink mode. We emphasise that even a small intensity perturbation 
would also point towards the fast kink MHD mode. However, one has to consider how the scales of the photospheric driver compare with the scales of individual threads. If the photospheric driver has dimensions smaller than the threads, the idea of kink waves is favourable. As we are dealing with waves in chromospheric and coronal structures, the kink interpretation becomes more meaningful if the structures are stable over several wave crossing times.

\begin{figure}
\includegraphics[width=0.99\textwidth]{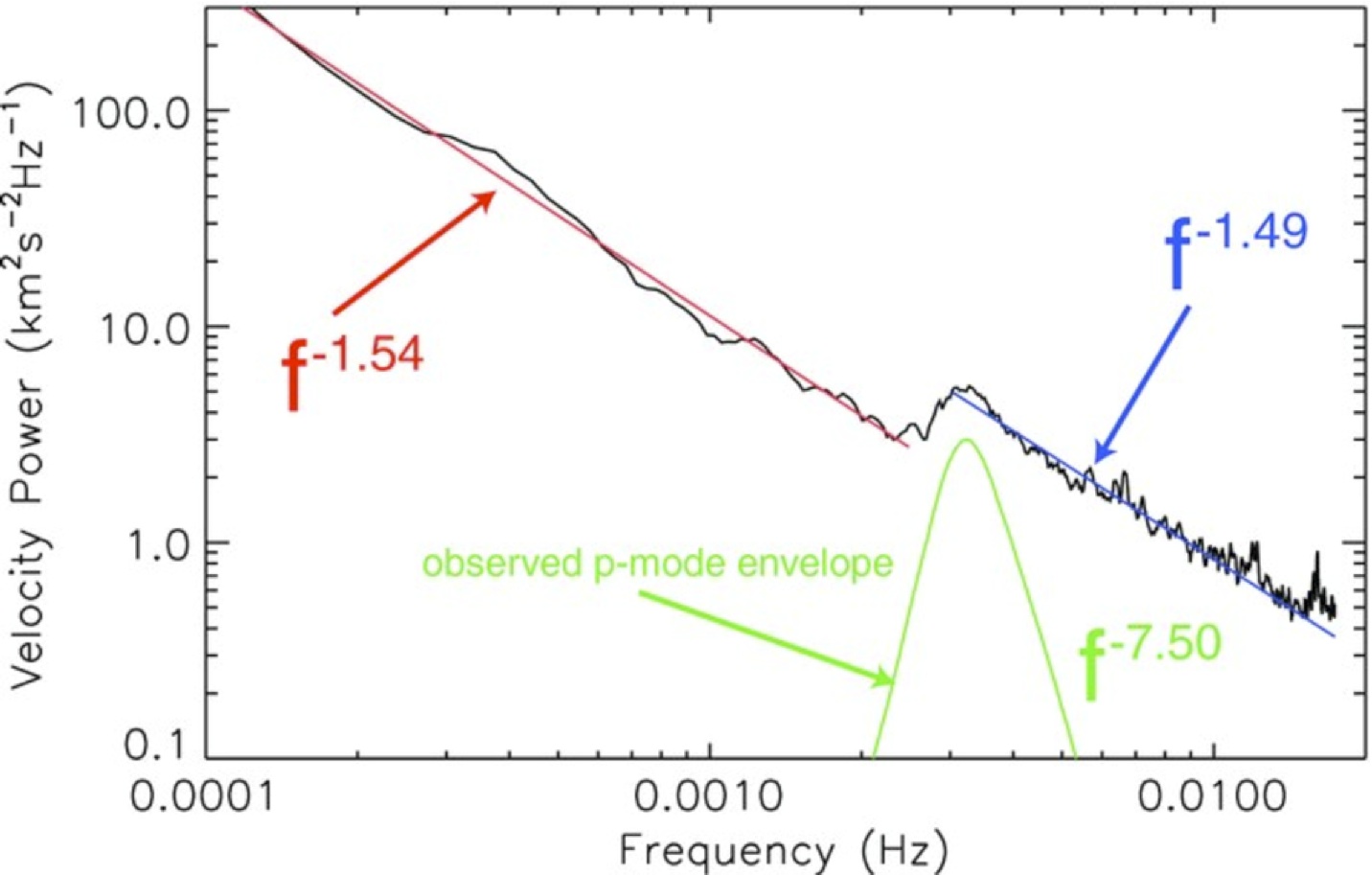}
\caption{The Fourier spectrum of the Doppler velocities in the corona (black line) 
and the photospheric $p$-mode spectrum (green line).  Power-law fits to the low 
(red) and high frequency (blue) parts of the coronal spectrum are also shown 
\citep[reproduced from][]{RefM}. }
\label{fig:2}       
\end{figure} 
 
Irrespective of the actual interpretation, there is no doubt that the observational findings of 
\cite{RefJ} are sound, and clearly demonstrate the ubiquitous nature of waves in the solar 
corona. In a follow up investigation, \cite{RefM} performed time-distance seismology of the 
solar corona, and constructed a {$k$--$\omega$} diagram of the region. The power spectrum 
of the coronal velocity perturbations shows an excellent agreement with that of the solar $p$-modes, 
which supports the suggestions that these waves can be transported into the corona 
along inclined flux tubes originating in the photosphere (Fig.~\ref{fig:2}). However, the 
high-frequency spectrum is distinctly different from the photospheric power spectrum, and 
indicates that the high-frequency component of these waves may be generated locally.
An estimate of the wave energy reveals that it is significantly smaller than the energy 
required to balance the coronal radiative losses. However, 
it should be emphasised that the low spatial resolution of CoMP, limited to 4.5 
arcseconds/pixel, overlooks small-scale solar structures, and therefore significantly 
underestimates the energy carried by these waves.

The higher spatial resolution of the Atmospheric Imaging Assembly on board the 
Solar Dynamics Observatory \citep[SDO;][]{Boe11} has indeed shown transverse 
oscillations in the corona with amplitudes of around 20~km{\,}s$^{-1}$. The 
amplitude of these motions is a factor of 40 higher that those found by 
CoMP, and may be sufficiently energetic to balance the radiative losses from the quiet 
solar corona \citep{RefAx}.

\subsection{Flare induced Alfv{\'{e}}n Waves}

Solar flares often induce temperature rises in the chromosphere of 
the order of 100 -- 200~K. This requires a significant energy input, amounting 
to 10 -- 20~erg{\,}cm$^{-3}${\,}s$^{-1}$ at these atmospheric heights. The dense nature 
of the lower solar atmosphere, when compared to the corona, makes it difficult 
to reconcile this amount of heating from an individual energy release in the 
corona \citep{RefN}. This is often called the ``number problem'', 
whereby the high total number of electrons required to be accelerated down into 
the lower atmosphere do not exist at coronal heights \citep{RefO}. 

An alternative interpretation to explain lower-atmospheric heating is that of 
flare-induced dissipation of Alfv{\'{e}}n waves. \cite{RefP} 
have suggested how Alfv{\'{e}}n-wave perturbations may be generated along 
current sheets, linking the corona to the lower solar atmosphere, in the aftermath 
of an impulsive flare event. As the group velocity of Alfv{\'{e}}n waves 
will be parallel to the magnetic field, the energy contained within the mode 
will remain trapped, providing the wavelength is small compared with the scale 
length of field and plasma variations. Thus, energy can be readily transported 
down into the lower solar atmosphere, where dissipation mechanisms convert the 
energy into localized heating.
The Poynting flux of the waves is proportional to the Alfv{\'e}n speed 
and the square of the magnetic field perturbation. X-class flare energy requirements 
would suggest a Poynting flux of approximately 10$^{11}$~erg{\,}cm$^{-2}${\,}s$^{-1}$.  
The magnetic field variations, as high as 200~Gauss \citep{sudol05}, observed 
during the most energetic solar flares, combined with the Alfv{\'e}n speed in the 
corona, implies that there is sufficient energy to meet the flare requirements \citep{RefO}.

\begin{figure}
\includegraphics[width=0.99\textwidth]{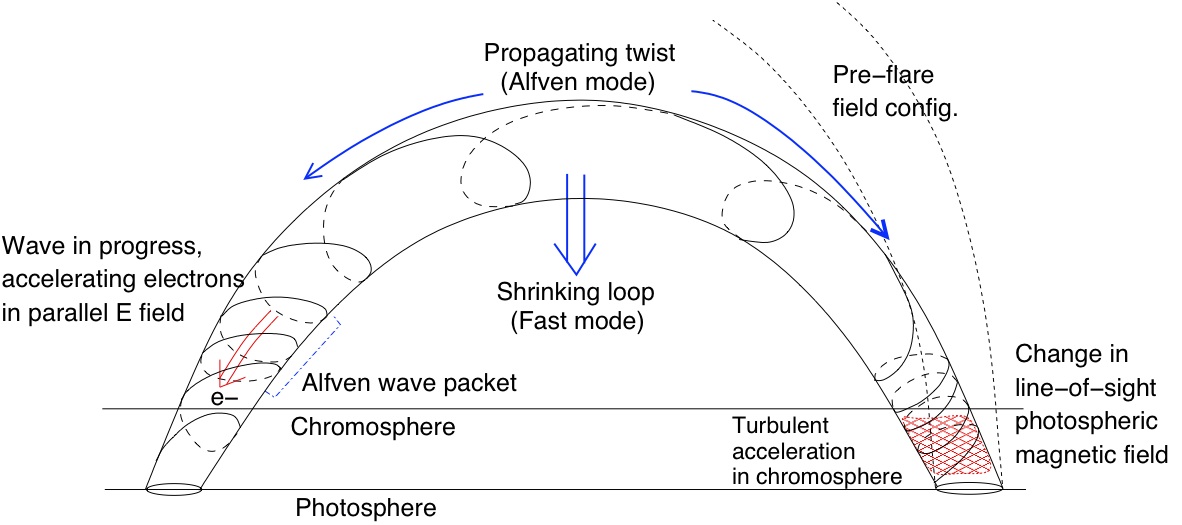}
\caption{The coronal magnetic field launches a torsional Alfv\'en wave pulse 
into the lower atmosphere, as well as a fast-mode wave pulse. That fraction of 
the Alfv\'en wave energy that survives into the chromosphere can also lead 
to stochastic acceleration there. The wave will be partially reflected from the 
steep gradients in the chromosphere (not shown) and re-enter the corona 
\citep[reproduced from][]{RefO}.}
\label{fig:3} 
\end{figure}

What triggers Alfv{\'{e}}n waves during flare events is still an open question. 
However, current understanding suggests they can be generated during the 
reconnection process, when the magnetic field relaxes from its stressed 
pre-flare state \citep{RefQ}. In this regime, magnetic field lines may still be 
highly distorted from a potential configuration, providing the necessary 
high-tension force required to induce Alfv{\'{e}}n waves \citep[Fig~\ref{fig:3};][]{RefO}.

A high degree of Alfv{\'{e}}n wave interaction will occur as these waves impact 
the lower solar atmosphere. \cite{RefR} have shown how incident 
Alfv{\'{e}}n waves will produce a fast-wave spectrum containing rapidly 
varying amplitudes. It is these waves which can be locally damped through 
ion-neutral or resistive damping, or form a turbulent 
cascade which propagates deeper into the photosphere \citep{RefS}.  

Magnetic reconnection can also produce small scale X-ray jets when a burst of 
solar plasma is driven into the corona at high velocities \citep{RefAs}. These X-ray 
jets, first observed by Yohkoh, have been studied in greater detail by the 
Hinode X-ray Telescope (XRT). The XRT studies identified a large number of 
X-ray jets moving with velocities in the range of 200 -- 800 km{\,}s$^{-1}$ \citep{RefAt}. 
The high end of these values is similar to the Alfv\'en velocity in the corona. 
The suggestion that an Alfv\'en wave may be generated in these reconnection 
sites, prompted \cite{RefAt} to search for transverse oscillations associated with 
the location of the jets. They identify transverse motions with a period of 
approximately 200~s, which seem to be related to the high velocity component 
of the outflow.

\section{The Photosphere and Chromosphere}

\subsection{Transverse Alfv\'en Waves in Spicules and Fibrils}

The wealth of observations from a wide range of ground-based and space-borne 
missions have shown that the solar chromosphere is permeated by spicules 
and fibrils. Spicules are dynamic, straw-like magnetic structures that are best identified 
in the Ca {\sc{ii}} H \& K lines, Na D$_3$, H${\alpha}$ and Ca {\sc{ii}} 8542~\AA, 
and can be divided into two distinct classes \citep{zaq-erd09}. Many detections of 
transversal waves in spicules have been tied to Type~{\sc{i}} spicules, which
are ubiquitous throughout the solar atmosphere, and longer lived ($\approx$10~min) than their
jet-like Type~{\sc{ii}} counterparts \citep{RefT}. However, the classification of spicules into two separate types has been recently called into doubt  \citep{zhang12}. 
Formed in low plasma $\beta$ regions, the axis of spicules and fibrils outlines the 
direction of the magnetic field. Transversal oscillation in these structures, interpreted as 
Alfv\'en waves, have recently been reported by \cite{RefT}. Transverse waves with flows of chromospheric material were also detected in low laying 
corona loops \citep{ofman08}.These authors used the 
Solar Optical Telescope \citep[SOT;][]{Tsu08} on Hinode to analyze a time 
series of Ca {\sc{ii}} H (3968~\AA) images. The Hinode observations clearly show 
that spicules undergo transverse motions, with periods in the range of 100 -- 500~s and 
amplitudes of 10 -- 25~km{\,}s$^{-1}$.  The oscillations are evident in traditional 
$x-t$ plots, where the intensity is plotted as a function of time in a direction perpendicular to the axis 
of the spicule. The estimated periods can be uncertain as they are comparable 
to the lifetime of the spicules. Estimates of the energy flux carried by these waves 
reveals that it is sufficient to balance the radiative losses from the corona and power the solar 
wind. Recently, \cite{RefU} used a high-cadence mode on Hinode to identify 
high-frequency oscillations in spicules with periodicities less than 50~s. These waves 
are interpreted in terms of upwardly propagating Alfv\'en 
waves with phase speeds $\geq$75~km{\,}s$^{-1}$, and wavelengths less than 
8~Mm \citep{RefMartinez}.  

\begin{figure}
\includegraphics[width=0.99\textwidth]{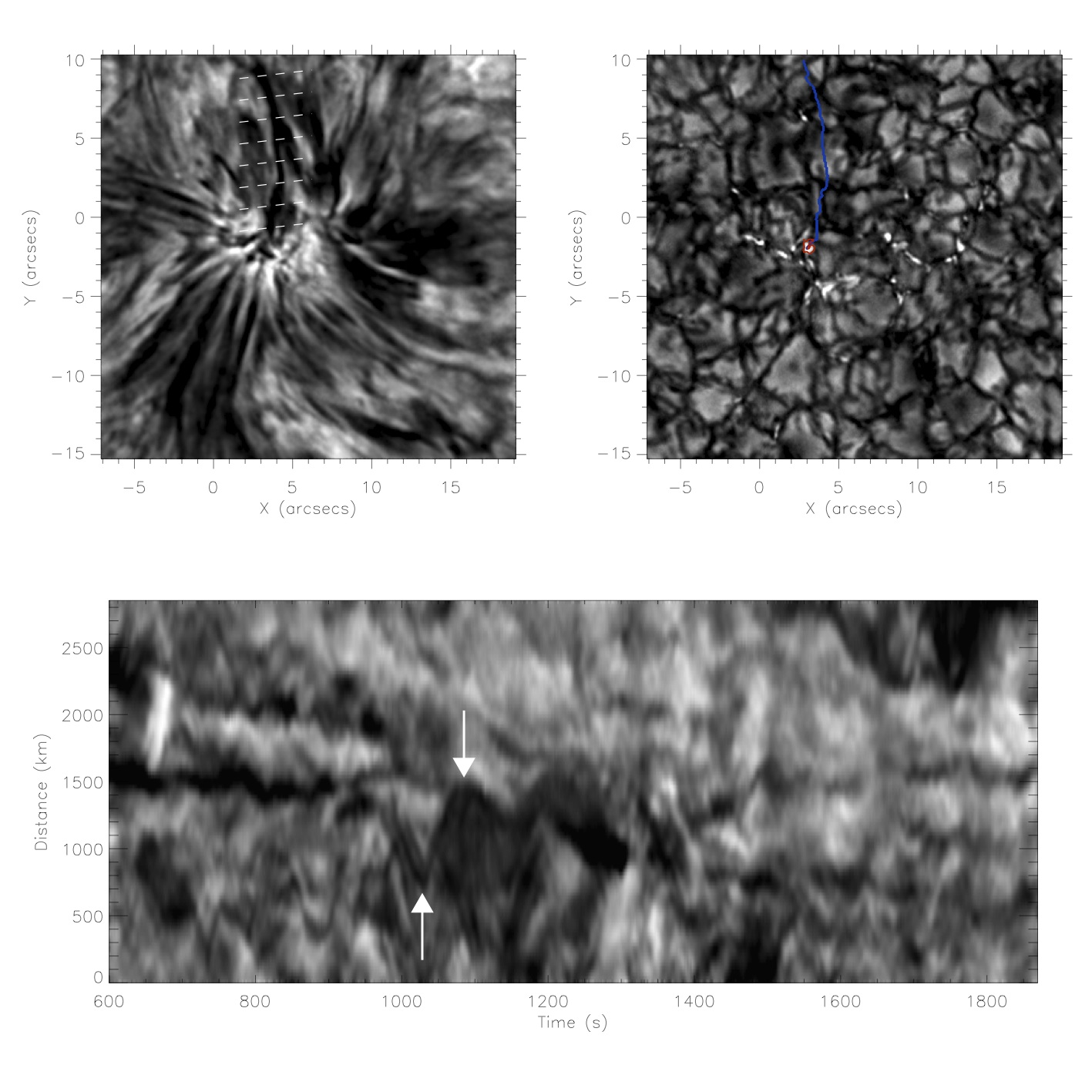}
\caption{Simultaneous images of the H$\alpha$ core (chromosphere; upper left) 
and G-band (photosphere; upper right), where dashed white lines in the 
H$\alpha$ core image highlight the spatial positions where 
$x-t$ cuts were made. The blue trace in the G-band image denotes the position of a
chromospheric spicule (visible in the left panel as a dark, straw-like structure), 
where one end is anchored into the photosphere above a MBP. 
A red contour indicates the location where a high concentration of longitudinal
oscillatory power is present. The lower panel is a sample H$\alpha$ time-distance cut, 
obtained 4000~km (5.5 arcsec) from the underlying MBP, revealing an abundance of periodic 
transverse motions in the solar chromosphere. White arrows highlight a trough and a 
peak of a typical transverse oscillation. The scale is in heliocentric coordinates, where 
1 arcsec approximately equals 725 km \citep[reproduced from][]{RefAv}.}
\label{fig:4} 
\end{figure}

While transverse kink motion has been shown to be abundant in spicule observations, 
the underlying cause of the periodic motions has remained speculative. 
Overshooting of convective motions in the photosphere, granular buffeting, 
rebound shocks, and global $p$-mode oscillations have all been suggested as 
candidates for the creation of spicule oscillations \citep{Rob79, Ste88, Vra08}. 
\cite{RefK} have suggested how photospheric $p$-modes can develop into 
shocks, and drive the dynamics of chromospheric spicules. The authors 
indicate how, even though low frequency $p$-modes become evanescent in the 
upper photosphere, these waves can propagate upwards into the 
upper chromosphere and corona through inclined flux tubes. 
Flux tubes with high inclination angles enhance the leakage of p-modes 
into the upper layers of the solar atmosphere. 
Recently, \cite{RefAv} have presented unambiguous evidence of how the generation of 
transverse oscillations in spicule structures is a 
direct result of the mode conversion of longitudinal waves in the lower 
solar atmosphere. Furthermore, these authors show how 
spicules are intrinsically linked to photospheric flux concentrations, with their 
spectrum undoubtedly including signals from both the 3-minute and 5-minute $p$-mode 
oscillations (Fig.~\ref{fig:4}).

The question of how these waves dissipate remains open. Wave energy damping 
in the inner corona can take place through, e.g. resonant absorption 
 \citep{ion78,davila87,holweg88,poedts89,poedts90,steinol93,ofman94,ofman95b,ofman95c,erd-goo95,rud-rob02,goo_etal11}, 
phase mixing \citep{hey-pri83,ofman02},  Alfv\'en turbulence 
\citep{RefV, vanBallegooijen} and/or mode conversion \citep{RefW}. 
\cite{velli89} suggested wave-reflection in the steep density gradients of interplanetary medium as a source of the turbulent cascade of unidirectional Alfv{\'e}n waves. This idea was applied to the lower solar atmosphere by \cite{matthaeus99} who proposed that the Alfv{\'e}n waves generated in the photosphere and chromosphere interact with their reflections to drive the effects of  turbulence offering a viable mechanism for heating the corona in the open field regions. Motivated by the CoMP observations, \cite{verdini09} have re-examined the turbulent spectrum from the base of the corona and out to 17 R$_\odot$. They find that turbulent dissipation can account for at least half of the heating required to sustain the solar wind. Reflection driven turbulence can therefore play a vital role in the acceleration of the fast solar wind. However, the main limitation of these models is the use of incompressible reduced MHD equations. Thus, the likely important coupling between Alfv\'en waves and compressive modes is neglected in these models.
\cite{RefU} suggest that 
Alfv\'en waves could instead be damped by cyclotron resonance in the 
heliosphere, out to 60~R$_{\odot}$, a mechanism that is also favourable for the heating of the solar wind (see reviews by \cite{marsch06,ofman10}).  

The evolution and dissipation of Alfv{\'e}nic perturbations in force-free magnetic structures 
has been studied in detail by \cite{malara00}, who examined the competing effects of phase 
mixing and three dimensional effects (i.e. exponential separation of field lines) in wave packet 
propagation. They find that in a chaotic magnetic field configuration, the dissipation is dominated 
by three dimensional effects, while in simpler equilibrium situations both phenomena 
contribute in spatially separated regions. We point out that this approach is limited by the 
assumption of a incompressibility. Large amplitude Alfv{\'e}n waves are subject to parametric 
decay as a result of their coupling with acoustic and transverse waves. The energy of an 
Alfv{\'e}n wave propagating in a turbulent medium can be transferred to other wave modes. 
\cite{malara96} have shown that the parametric instability of nonmonochromomatic waves 
decreases quadratically with departures from the monochromatic case. Numerical simulations 
of the time evolution of large amplitude Alfv{\'e}n waves have confirmed the development 
of this instability in the solar wind \citep{malara00a, delzanna01}.

Spicule oscillations may also be attributed to the transversal kink MHD mode \citep{zaq-erd09}. 
The observational signatures of the guided-kink magneto-acoustic mode are the 
swaying, transversal motions of magnetic flux tubes, similar to the oscillatory 
behaviour observed in spicules \citep{RefX}. \cite{RefT} have argued that, 
due to their short lifetime, the spicules cannot act as stable wave guides and 
the kink mode cannot satisfactory explain the Hinode observations. 
This apparent contradiction may be resolved by considering the spicule as 
sliding along the oscillating field lines. 
 
In view of the recent Hinode SOT analysis of spicules, \cite{RefH} have 
re-examined the SUMER results of transition region line widths by combining 
the results of forward modelling with spicule observations. The authors conclude 
that the C~{\sc{iv}} intensities and non-thermal velocities can be explained by the 
superposition of longitudinal mass flows and ``Alfv\'enic motions''. It remains unclear what 
is meant by ``Alfv\'enic motions'', since the eigensolutions of MHD equations are not 
labelled commonly as such in theoretical studies. On the other hand it is not evident that 
linear MHD eigenmodes of cylindrical 
magnetic structures can well-describe all realistic coronal conditions.
The apparent isotropy in the center-to-limb variations 
may be explained by the similar magnitude of observed longitudinal and 
transverse velocity components.

Solar prominences, the heavy and cool elongated magnetic structures supported 
by mainly horizontal magnetic fields in the solar corona, can support a wide 
variety of MHD waves. Analysing the properties of these waves can give an 
unprecedented insight into the small-scale structuring of these features, with clues 
about the magnetic field strength through means of solar magneto-seismology \citep{nakariakov05}. 

Prominence oscillations cover a wide range of amplitudes.
The large amplitude velocity oscillations, $>$20~km{\,}s$^{-1}$, are thought to be excited by 
Moreton waves produced by flares \citep{moreton60}. These oscillations are observed 
in both the transverse and longitudinal directions, with periods in the range of tens of 
minutes to a few hours \citep{tripathi09}.  The potential of large-amplitude prominence 
oscillations for deriving the vertical magnetic field was first explored by \cite{hyder66}, 
who obtained field strengths of tens of Gauss. More recently, \cite{vrsnak07} used 
prominence seismology techniques to infer the Alfv{\'e}n speed, as well as the azimuthal 
and axial components of the magnetic field strength. The small amplitude oscillations, 
$<$3~km{\,}s$^{-1}$, are usually restricted to only part of the prominence, and are 
thought to be excited by the photospheric and chromospheric oscillations.

The vertical oscillatory motions observed in prominences have been interpreted 
as transverse waves occurring on the horizontal magnetic field lines of these structures. 
The oscillations have periods of 2 -- 4~min, and are in phase along the entire 
length of the prominence thread, as observed in the plane of the sky \citep{RefAr}. 
The lack of any additional information make the exact wave mode difficult to 
disentangle. We therefore believe that the suggestion that these oscillations are 
due to Alfv\'en waves may be somewhat premature. This is further supported 
by numerical simulations by e.g. \cite{erd-fed07}. 
For a recent review on prominence oscillations see \cite{arregui12}.

\subsection{Torsional Alfv\'en Waves}

The periodic motions of photospheric footpoints, in an axially symmetric 
system, can excite torsional ($m=0$) Alfv\'en waves \citep[see \S\ref{sec:theory} 
of this review and][]{RefY,rud99}. 
The observational signature of torsional Alfv\'en waves, with a velocity
component along the observers line-of-sight, will arise from torsional 
velocities on small spatial scales. It 
is suggested that these torsional velocities will produce an observable 
full-width half-maximum (FWHM) oscillation in a given line profile. 
At the linear limit of infinitesimally small amplitude, Alfv\'en waves are 
incompressible, and should therefore 
exhibit no periodic intensity variations. 

The detection of such torsional motions in the lower solar 
atmosphere has, until recently, remained elusive. \cite{RefZ} used the Swedish Solar 
Telescope, equipped with the Solar Optical Universal Polarimeter 
\citep[SOUP;][]{Tit84}, to carry out imaging spectroscopy of the lower solar atmosphere. 
The tuneable nature of SOUP allowed complete sampling of the H$\alpha$ profile, and 
multi-object multi-frame blind deconvolution 
\citep{RefAa} image reconstruction was implemented to remove small-scale 
atmospheric fluctuations. 

\begin{figure}
\includegraphics[width=0.99\textwidth]{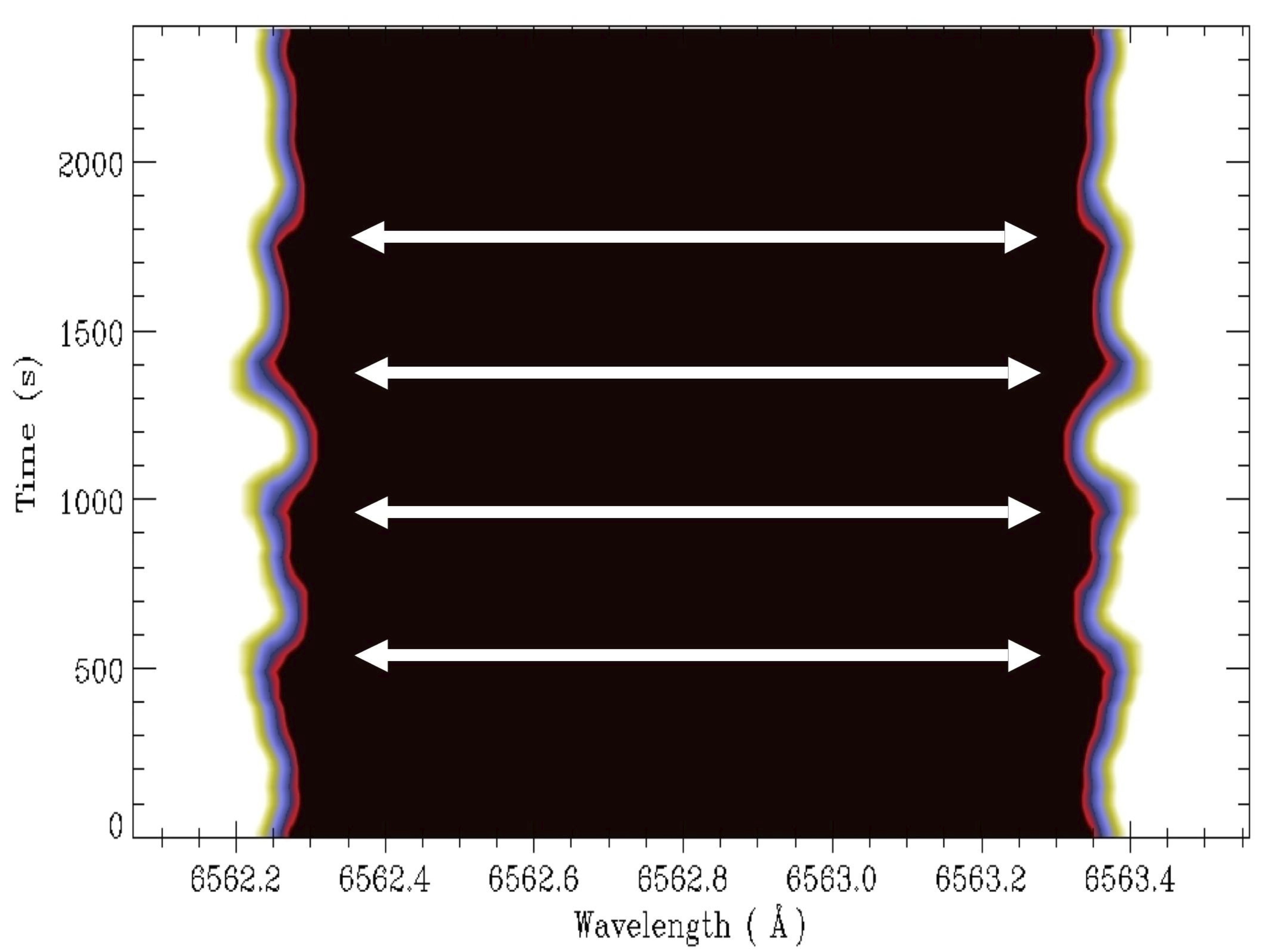}
\caption{A wavelength versus time plot of the H$\alpha$ profile showing the variation of 
FWHM line width as a function of time. The arrows indicate the positions of maximum 
amplitude of a 420-s periodicity associated with the MBP group 
\citep[reproduced from][]{RefZ}.}
\label{fig:5} 
\end{figure}

\begin{figure}
\includegraphics[width=0.99\textwidth]{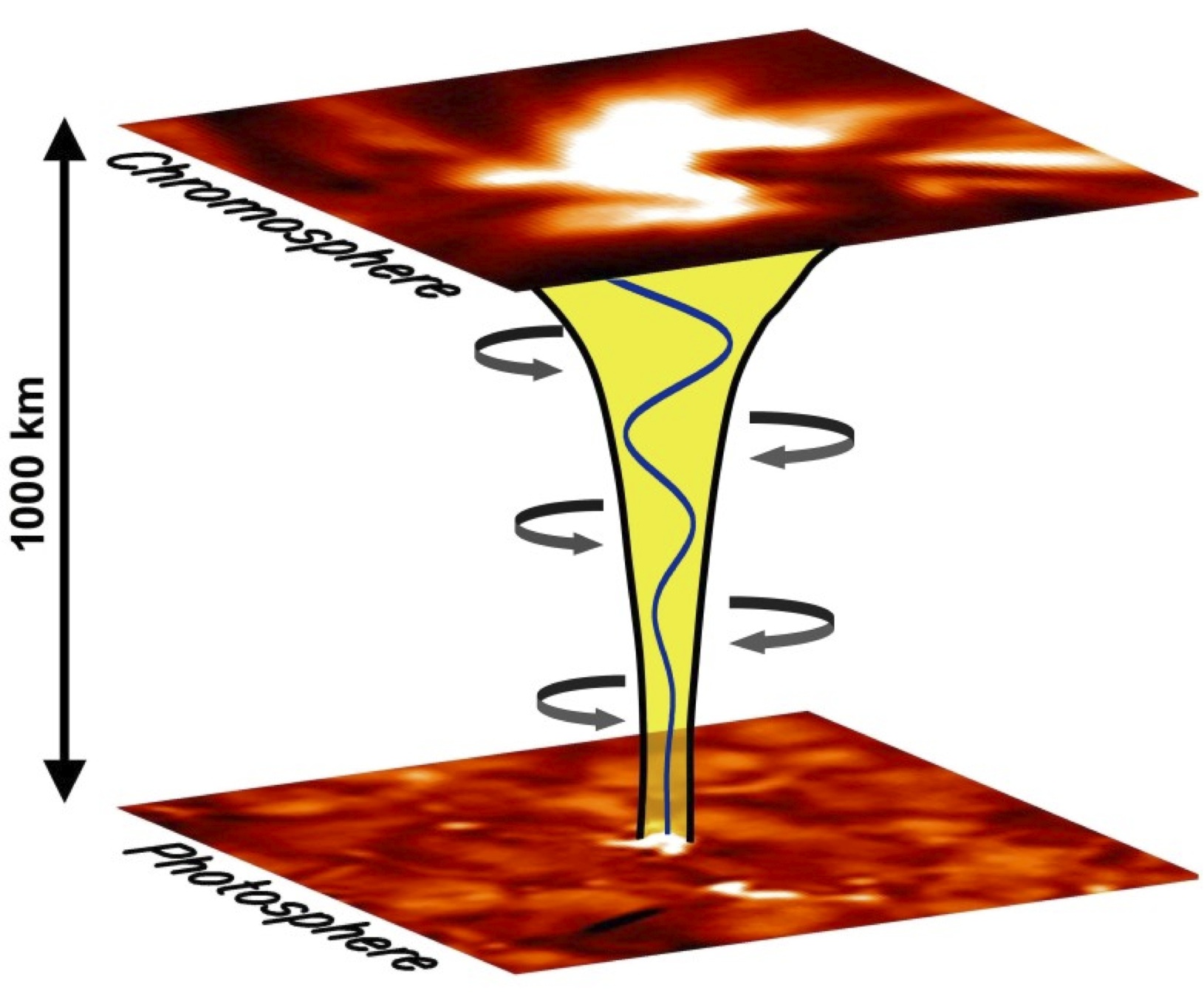}
\caption{Schematic diagram of an expanding magnetic flux tube, sandwiched 
between photospheric and chromospheric intensity images, undergoing a 
torsional Alfv\'en perturbation that propagates in the vertical direction. At a 
given position along the flux tube, the Alfv\'en displacements are torsional 
oscillations that remain perpendicular to the direction of propagation 
\citep[not to scale; reproduced from][]{RefZ}.  {A divergent magnetic geometry leads to wave amplitude increase with height and coupling to compressional waves \citep{ofman98}.}}
\label{fig:6} 
\end{figure}

The authors investigated a large conglomeration of photospheric 
magnetic bright points (MBPs), covering an area of 430,000~km$^{2}$. 
FWHM oscillations of the 
H$\alpha$ line profile, in addition to line-of-sight Doppler blue-shift velocities 
of 23~km{\,}s$^{-1}$, provided evidence for the presence of torsional (m=0) 
Alfv\'en waves. The progression of the Doppler-compensated 
H$\alpha$ FWHM is shown as a function of time 
in Figure~\ref{fig:5}, with a schematic diagram of the observed wave mode 
shown in Figure~\ref{fig:6}. A lack of 
periodic intensity fluctuations in their dataset further strengthened the 
authors conclusions.

\begin{figure}
\includegraphics[width=0.99\textwidth]{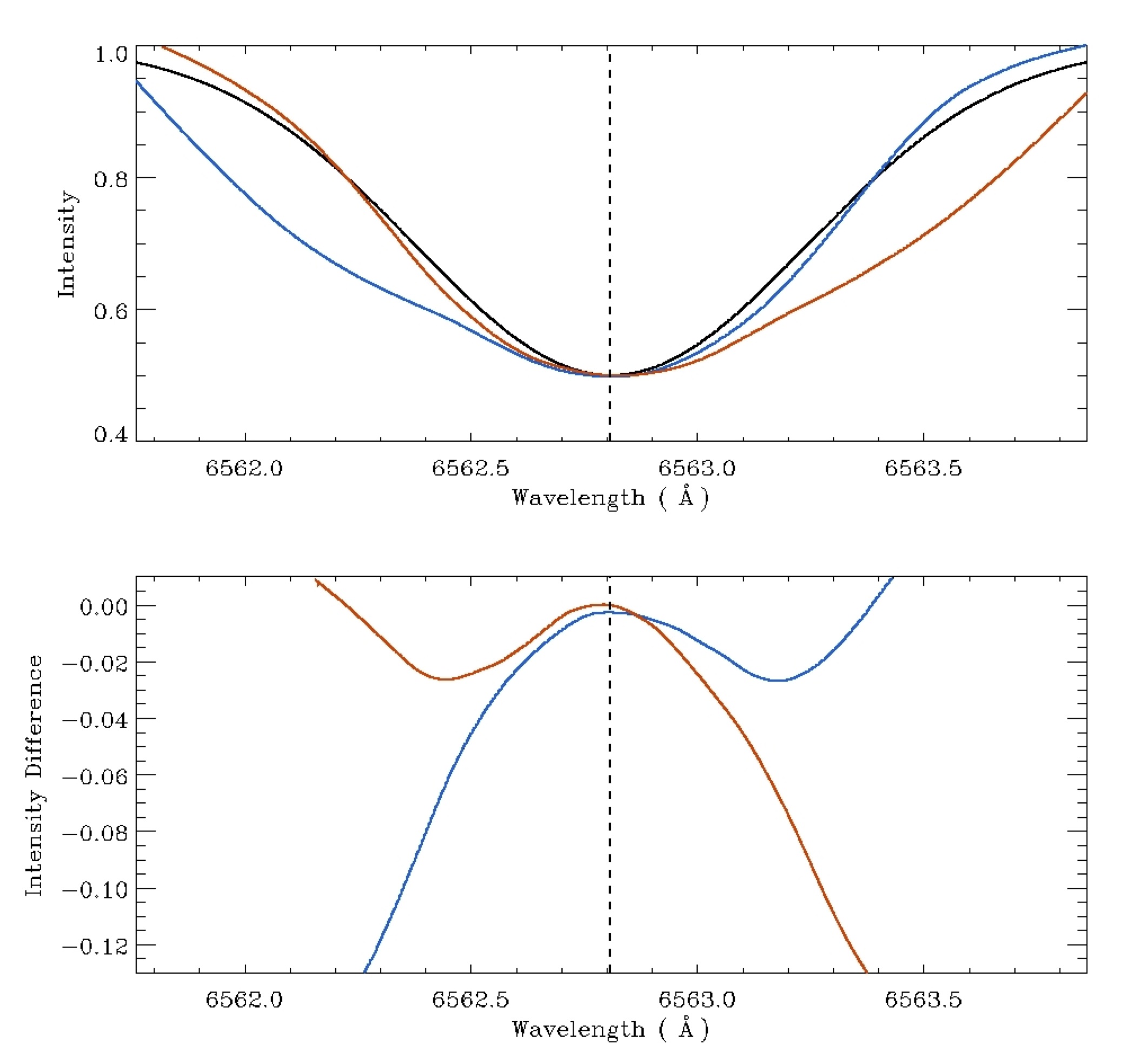}
\caption{{\it{Top Panel:}} Simultaneous plots of the H$\alpha$ line profile 
at opposite edges of the MBP group. The red line shows a profile which 
is heavily broadened in the red wing of the H$\alpha$ profile, while the 
blue line displays a large degree of non-thermal broadening in the 
blue wing of the H$\alpha$ profile, at the opposite edge of the MBP group. 
A rest H$\alpha$ profile is displayed as a solid black line, with the H$\alpha$ 
core rest wavelength plotted as a vertical dashed black line. 
{\it{Lower Panel:}} The same broadened profiles 
plotted above, but now displayed as intensities relative to the rest 
H$\alpha$ profile. It can be seen that the resulting profiles are 
anti-symmetric about the H$\alpha$ core rest wavelength of 6562.808{\AA}, as 
indicated by the vertical dashed black line. This out-of-phase phenomena, 
at opposite edges of a flux tube undergoing torsional displacements, is 
a characteristic of the m=0 Alfv\'en mode.}
\label{fig:alfven} 
\end{figure}

 \cite{RefAc} have used the Interferometric Bi-dimensional Spectrometer 
\citep[IBIS;][]{Cav06} to perform imaging spectroscopy of the solar chromospheric 
network in the H$\alpha$ and Ca {\sc{ii}} 8542~\AA\ line. They find that an 
increased H$\alpha$ line width correlates well with the Ca {\sc{ii}} minimum 
intensity. This is interpreted as evidence for shock heating in the magnetic 
network which raises the possibility that acoustic shocks can introduce an 
oscillatory behaviour in the observed H$\alpha$ line width. 
To investigate this possibility further, \cite{RefZ} carried out a 
number of additional tests to help verify or refute their original conclusions. 
Since torsional Alfv\'en waves will produce a FWHM oscillation that is
180$^{\circ}$ out of phase at opposite boundaries of the waveguide \citep{RefL}, 
the authors investigated the instantaneous H$\alpha$ profile at opposite edges of 
the MBP group. Through comparison to an at-rest profile, the nature of 
wing broadening is clearly visible in the lower panel of Figure~\ref{fig:alfven}. 
These findings reinforce the conclusion of \cite{RefZ} that the observed FWHM 
oscillations are due to the presence of torsional Alfv\'en waves.

The energy flux carried by the Alfv\'en waves can be estimated using the WKB approximation, 
\begin{equation}
F_A = \rho v^{2} v_{A} \ ,
\end{equation}
where $\rho$ is the mass density of the flux-tube, $v$ is the observed 
velocity amplitude, and $v_A$ the Alfv\'en speed. For a mass density of 
$\rho \approx 1 \times 10^{-6}$~kg m$^{-3}$ \citep{RefAe}, and utilizing the 
derived wave parameters described by \cite{RefZ}, the localised energy flux in 
the chromosphere is $\approx$150,000~W{\,}m$^{-2}$. It is estimated that, at any
given time, $\approx$1.6\% of the solar surface is covered by bright 
point groups \citep{San10}, producing a global average of 2400~W{\,}m$^{-2}$. 
This value is significantly above the energy required to heat the corona and 
power the solar wind.
We emphasise that the above energy estimate is correct in the absence of 
vertical gradients or when they can be neglected in the WKB limit. The presence of a soalr wind enhances the 
transmission of low frequency waves, while high frequency waves 
have a transmission similar to static models \citep{heinemann80,velli93}.

\section{Wave Drivers}

Turbulent convection leads to the stochastic excitation of $p$-mode oscillations, 
which are ubiquitous in the lower solar photosphere, and can be detected in both 
spectral line intensities and Doppler shifts. The frequency of $p$-mode oscillations 
are in excess of 1~mHz, with the strongest power detected at $\approx$3~mHz 
(5 min). However, the $p$-modes can be modified by the magnetic field 
\citep[for reviews with references on this subject see e.g.][]{erd06a, pin-erd11}. In magnetic 
regions such as sunspots and pores, the amplitudes of the 5-min 
oscillations are suppressed, while high frequency oscillations, above the 
acoustic cut-off, have enhanced power. The suppression of 
low frequency oscillation occurs on very small scales (a few arcseconds), 
while the enhanced high frequency power occurs 
primarily in patchy halo-like areas surrounding the strongest magnetic fields 
\citep{RefAg}. 

The solar magnetic network, defined by the boundaries of super-granular flows, 
shows evidence for MHD waves with frequencies in the range of 1 -- 4~mHz. 
The internetwork has frequencies in excess of 4~mHz, which have been 
modelled as acoustic waves driven by photospheric velocities \citep{RefAh}. 
Network bright points are formed at the boundaries of super-granules, and 
show a strong correspondence with the underlying magnetic field. The granular 
buffeting of photospheric flux tubes can create transverse waves which 
propagate along the field lines and into the upper atmosphere. The models of 
\cite{RefAi} suggest that the low frequency oscillations in the magnetic network 
are due to transverse waves that are generated by the granular buffeting of 
photospheric flux tubes. Density stratification will lead to an increase in the 
wave speed as they travel into the upper photosphere and chromosphere, where 
mode transformation can occur. Transverse waves can transfer power to 
longitudinal modes, which then develop into shocks and heat the surrounding 
plasma. A number of criteria may be used to identify if mode transformation has 
occurred within a given network bright point \citep{RefAj, RefAk, Vigeesh}. 
These include:
\begin{itemize}
\item{Power at a particular frequency, $\nu$, as identified in the 
upper-photosphere/lower-chromosphere, should also be identified in the upper chromosphere, 
but with reduced amplitude.}
\item{Power at double frequency, 2$\nu$, should also appear in the upper chromosphere.}
\item{Waves remaining at, or above the transverse cut-off, should propagate at 
approximately the sound speed.}
\item{The oscillatory signal should be quasi-periodic, as a result of the granular buffeting.}
\end{itemize}
The studies of \cite{RefAl} and \cite{RefAm} have provided the observational 
evidence for mode coupling in such magnetic networks. In these examples, the 
authors provide evidence for transverse waves with frequencies in the range 
1 -- 2.5~mHz, transferring power to longitudinal waves with frequencies of 
2 -- 5~mHz. We emphasise that the above observational signatures could also appear 
as a result of nonlinearity.

It has been demonstrated, both analytically and theoretically, that 
when the magnetic pressure is approximately equal to the gas pressure 
(plasma $\beta=1$) in a solar flux tube, longitudinal to transverse mode 
coupling may also occur \citep{DeM04}. Until recently, 
this form of ``reversed mode coupling'' had not been verified 
observationally. However, \cite{RefAv} have suggested that 
longitudinal-to-transverse mode conversion may account for the transversal 
waves which are ubiquitous in the solar chromosphere. 
\cite{RefAv} utilised high spatial and temporal resolution observations, obtained 
with the Rapid Oscillations in the Solar Atmosphere \citep[ROSA;][]{Jess10} imager 
at the Dunn Solar Telescope, to investigate the origin of transverse oscillations in 
chromospheric spicules. The authors present strong evidence that transverse 
oscillations in Type {\sc{i}} spicules, with periods in the range of 65 -- 220~s, 
originate in longitudinal oscillations of photospheric magnetic bright points, at 
twice the transverse period. It must be emphasised that this form of mode 
coupling occurs in the reverse direction of that reported by \cite{RefAl}, 
i.e. longitudinal to transverse. The mechanism for the mode transformation 
revolves around the nature of the photospheric driver, whereby a 90$^{\circ}$ phase 
shift across the body of the spicule creates gradients in pressure, which displace 
the spicule axis, generating the observed kink modes. These periodic pressure 
differences also induce a frequency doubling of the coupled wave mode, in 
addition to producing periodic compressions and rarefactions along the body of the 
spicule (i.e. the sausage mode). 
 
\begin{figure}
\includegraphics[width=0.99\textwidth]{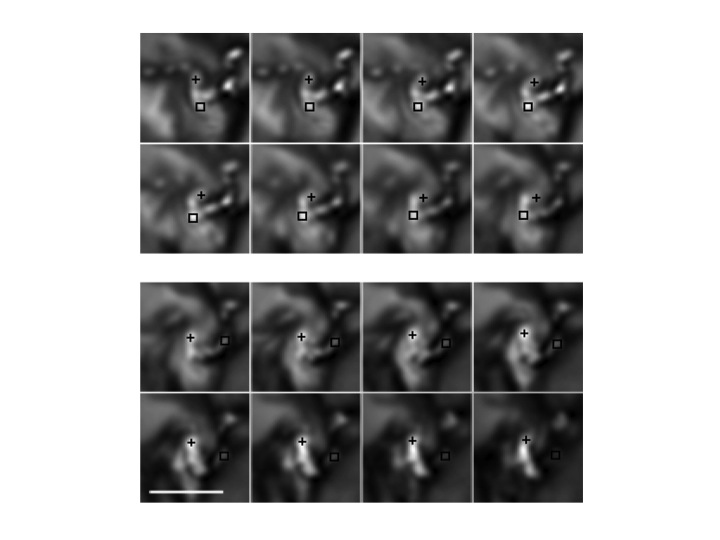}
\caption{A sequence of G-band snapshots showing  a photospheric vortex. 
Two pairs of MBPs are marked as squares and crosses to help the reader identify the rotational motions. The snapshots have been taken with a cadence of 
15 s. The bar in the lower left corner corresponds to 1,000 km on the solar surface 
\citep[reproduced from][]{RefAo}. }
\label{fig:8} 
\end{figure}

It is known that convective motions, intrinsic to photospheric granulation, 
transport angular momentum and 
agitate plasma in the outer layers of the Sun's atmosphere. Solar material rises 
in the granules and sinks in the inter-granular lanes. The plasma that returns 
to the solar interior has increased angular momentum, and therefore generates 
an increased amount of vorticity at the edges of the lanes \citep{RefAn}. The idea 
of vortices as the source of torsional wave modes 
has been proposed by \cite{hol82} and \cite{velli99}. Small scale 
vortex motions, with dimensions less than $\approx$500~km, have recently 
been observed in photospheric G-band images of MBPs 
\citep[Fig.~\ref{fig:8};][]{RefAo}. These small scale vortices have lifetimes 
of a few minutes, and appear to trace the pattern of a super-granular cell. 
They seem to be formed when two MBPs move towards a common point, 
and rotate relative to one another. \cite{RefAp} have observed similar vortices 
in the chromospheric Ca {\sc{ii}} 8542~\AA\ line core. The location of the 
chromospheric vortices coincides with photospheric MBPs, although there is 
no current evidence for simultaneous swirl events in the corresponding 
photospheric locations. 

Recent radiative MHD simulations have shown photospheric vorticities 
in intergranular magnetic field concentrations \citep{shel11}. The connection 
between MBP vortices and upper atmospheric swirls appears to be magnetic in nature. 
Simulated G-band images show a direct comparison between magnetic vortices and rotary motions of photospheric bright points. 
These swirly motions of plasma in the intergranular magnetic field concentrations could be 
responsible for the generation of different types of MHD wave modes such as kink, sausage 
and torsional Alfv\'en waves.

 
\section{Theory}
\label{sec:theory}
 
\subsection{Alfv\'en Waves in Uniformly Magnetised Plasma}
\label{subsec:aw_uni}

Within the linear limit of infinitesimally small wave amplitudes in 
uniform, finite beta magnetized plasma, there are three distinct types 
of MHD waves that can be mathematically described as: 
slow and fast magneto-acoustic waves, and Alfv\'en waves. The first two 
types of wave have an acoustic character modified by the magnetic field, 
whereas the Alfv\'en waves exist purely because of the presence of a magnetic
field. The three physically distinct MHD wave modes exist due to the following 
reasons. The MHD description of a magnetised plasma is
using macroscopic quantities, let's say density ($\rho$), velocity 
(${\bf v}$), magnetic field (${\bf B}$) and a thermodynamics variable, 
say pressure ($p$). In ideal MHD (i.e. no dissipation or non-adiabatic 
processes) these quantities are coupled through a set of eight non-linear partial
differential equations (PDEs) governing mass conservation, momentum 
conservation, governing magnetic flux and the energy conservation. If this is a
closed system, in a stationary state one may expect eight eigenvalues 
corresponding to eight different physical wave modes. However, there is
the solenoidal condition, that will reduce the number of independent 
PDEs to seven, resulting in seven eigenvalues. One eigenvalue, $\omega\equiv 0$, turns
out to be describing the entropy wave that does not carry information, and 
may be disregarded, leaving six possible independent eigenvalues
corresponding to six further independent modes. An interesting intrinsic feature 
of the remaining eigenvalues is that they appear squared, i.e. there is no
distinction between forward and backward propagating MHD waves in a 
{\it static} (i.e. ${\bf v_0}=0$) magnetised plasma. Hence, there are three
physically distinct MHD waves. These waves are a complete set and their 
appropriate linear superposition can be combined into any linear MHD
wave perturbations. This also means that the individual eigenmodes (slow, 
fast magneto-acoustic and Alfv\'en waves) are orthogonal to each
other and there is no linear transformation that can transform one MHD 
eigenmode into the other. This latter property is very important
when identifying MHD modes in observations. Often, an accurate mode 
identification is possible only by considering very carefully the mode dispersion or 
the phase relation between the perturbations involved. An interesting and 
educational example of this approach is e.g. by \cite{fuj_tsu09}. The problem 
becomes even more subtle in structured plasmas (e.g. magnetic cylinder or 
slab), where the geometry of the problem introduces an infinite set of MHD eigenmodes. 

\subsection{Alfv\'en Waves in Structured, Non-uniform, Magnetised Plasma}
\label{subsec:aw_smg}

\begin{figure}
\includegraphics[scale=0.25]{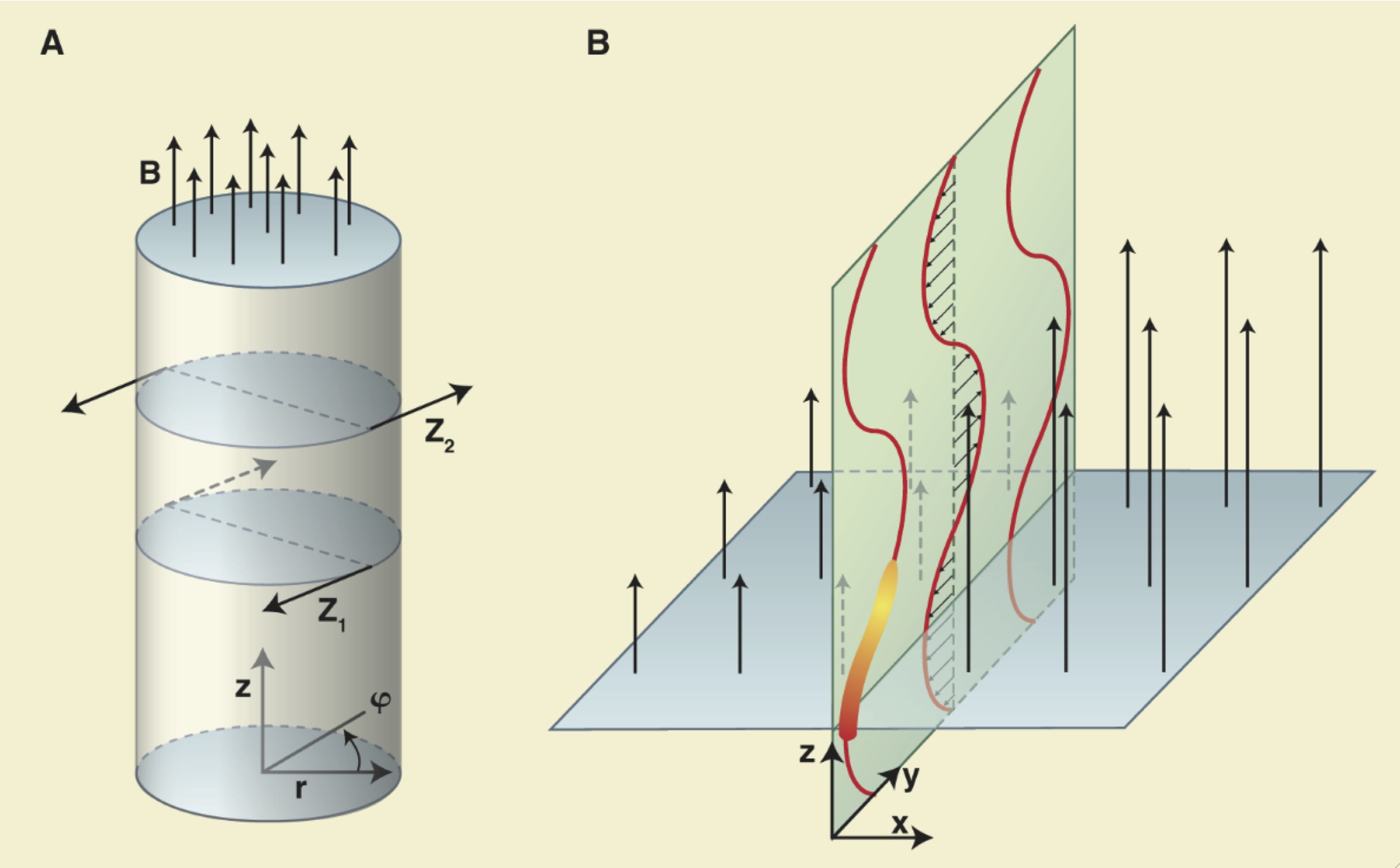}
\caption{(A) Magnetic flux tube showing two snapshots 
(at positions $z_1$ and $z_2$) of Alfv\'en wave perturbations propagating 
in the longitudinal $z$ direction along field lines.  At a given position the 
Alfv\'en perturbations are torsional oscillations (i.e. oscillations are in the 
$\varphi$-direction, perpendicular to the background field).
(B) Alfv\'en waves propagating along a constant magnetic surface. 
The Alfv\'en perturbations are within the magnetic surface ($yz$ plane) at 
the discontinuity, perpendicular to the background field ($y$ direction), 
whereas the waves themselves propagate along the field lines ($z$ direction). 
Density enhancements (e.g. in the form of spicules as observed 
by \citealt{RefT}, or within prominences as seen by 
\citealt{RefAr}) are visualised here as a yellow-red thin blob 
that follows the field lines. Vertical arrows indicate the magnetic field 
gradient increasing from left to right \citep[reproduced from][]{RefX}.}
\label{fig:re_aw} 
\end{figure}

A natural way to improve the simplistic view of a uniform MHD plasma 
(c.f. \S\ref{subsec:aw_uni}), in order to narrow the gap between theory and the 
observed wave processes taking place in the solar atmosphere, is to add 
non-uniformity. One way to achieve this goal is to introduce structuring. 
This may be done in many ways, but both the slab and cylindrical geometry 
seem to be a popular approach to model the building blocks of the solar 
atmosphere. We emphasize that the conservation of distinct modes 
in the real coronal plasma is not guaranteed, since the coronal plasma 
is nearly always more complex in nature than a collection of slabs 
and cylinders \citep{ofman09}. Another possibility to introduce non-uniformity is to consider 
inhomogeneity. Inhomogeneity may be either along the waveguide (for 
example for a vertical magnetic waveguide this could be the gravitational 
or magnetic stratification) or across (e.g. for a vertical flux tube, this may 
mean inhomogeneity in the radial direction). It was always expected that 
once MHD waves are generated, they will easily propagate along magnetic 
flux tubes (homogeneous or non-uniform), the building block of the solar 
atmosphere (Fig.~\ref{fig:re_aw}a), or along magnetic field lines at 
constant magnetic surfaces (Fig.~\ref{fig:re_aw}b). 

MHD wave theory has been around for a while, in particular for linear 
magneto-acoustic and Alfv\'en waves in structured magnetised plasmas 
(cf., \citealt{rob81,edw_rob82} for slab and \citealt{edw_rob83} for cylindrical 
geometries). The propagation of torsional Alfv\'en waves in vertical magnetic 
tubes have been studied by, e.g., \cite{hol81,hol82} and more recently in the 
context of filtering torsional Alfv\'en waves by \cite{fed_etal11b}. In order to overview 
briefly our current understanding of what kind of linear MHD waves are 
supported by MHD structures, we will assume a simple cylindrical geometry 
with some non-uniform radial inhomogeneity. We would also point out that it 
would not be practical to single out the Alfv\'en wave mode, as there is 
coupling between the linear MHD modes. The waves have to be investigated 
in a co-existing context. In nature, however, there may be a physical 
situation where the excitation mechanism (or driver) would favour one 
of the modes. Moreover, other wave modes will be produced due to linear 
(in non-uniform media) and non-linear (even in an initial inhomogeneous plasma) 
coupling. In order to have the problem tractable for the purpose of 
wave mode identification, with a focus on Alfv\'en wave, we assume a 
uniform magnetic twist here and will follow the key steps of \cite{erd-fed10}. 

\begin{figure}
\centerline{
\includegraphics[scale=0.55]{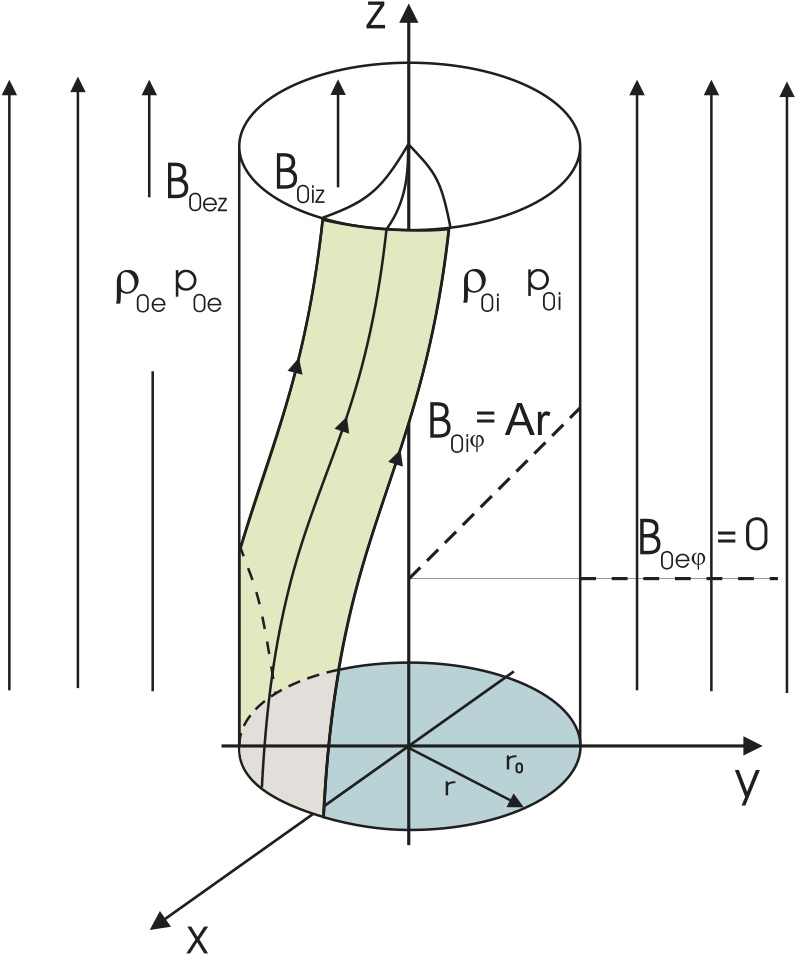}
}
\caption{The equilibrium configuration.  The uniformly twisted 
magnetic flux tube is in an ambient straight and uniform magnetic field. At the boundary of 
the flux tube there is a jump in magnetic twist \citep[reproduced from][]{erd-fed10}.}
\label{fig:re_twisted-cyl}
\end{figure}

Let us consider a magnetically twisted straight flux tube embedded within a uniformly 
magnetised plasma environment in cylindrical geometry. We note, such structures are always unstable, 
unless they are supported by appropriate boundaries and the twist does not exceed the critical twist for the MHD kink instability (e.g. \cite{raadu72}). 
The $z$-axis of the cylindrical 
coordinate system $(r, \varphi, z)$ is along the tube. The radius of the tube ($r_{\rm 0}$) is constant. 
The unperturbed state and geometry of the model of a magnetically twisted tube is sketched 
in Figure~\ref{fig:re_twisted-cyl}. All dependent variables inside the flux tube have 
index $i$, while quantities outside the tube are denoted with index $e$.
In cylindrical equilibrium the magnetic field and plasma pressure satisfy the equilibrium 
condition in the radial direction:
\begin{eqnarray}
\frac{\rm d}{{\rm d}r}\left(p_{\rm {0i}}+\frac{B_{{\rm 0i}\varphi}^2+B_{{\rm
0iz}}^2}{2\mu_0}\right) +\frac{B_{\rm 0i\varphi}^2}{\mu_0{r}}=0.
\label{re_eq.11}
\end{eqnarray}
Here, the second term in the brackets represents magnetic pressure and the third term of 
Equation~(\ref{re_eq.11}) derives from magnetic tension due to the azimuthal component of the 
equilibrium magnetic field, $B_{\rm 0i\varphi}$. Gravity is not 
included in the present analysis. The particular case of a uniformly twisted equilibrium 
magnetic field, $\bf {B_{0}(r)}$, of the form:
\begin{eqnarray}
\bf{B}_{0}(r) = \left\{ 
\begin{array}{ll}
 (0,Ar,B_{\rm {0iz}}), &\mbox{ $r\leq r_0$},   \\
 (0,0,B_{\rm {0ez})}, &\mbox{ $r>r_{\rm 0}$},
 \end{array} \right. \label{re_eq.12}
\end{eqnarray}
is considered, where
\begin{equation}
B_{{\rm 0iz}}=B\left(1-2\frac{A^2r^2}{B^2}\right)^{1/2} \nonumber
\end{equation}
in order to satisfy the pressure equilibrium governed by Equation (\ref{re_eq.11}). Here $A$ and 
$B$ are arbitrary constants. This choice of magnetic field may represent solar atmospheric 
flux tubes with observed weakly twisted field components \citep[see e.g.][]{kli00}. 

In a cylindrical coordinate system with the equilibrium state inhomogeneous in the
radial direction only, ${\bfxi}=(\xi_{\rm r},\xi_{\rm\varphi},\xi_{\rm z})$ is the 
Lagrangian displacement vector denoting perturbations from the equilibrium position; 
the unperturbed mass density 
$\rho_{0}$ and pressure $p_{0}$ are constants (though not equal to zero) inside the 
flux tube, respectively; ${\bf{B_{0}}}=(0,B_{0\varphi}(r),B_{\rm {0z}}(r))$ is the equilibrium 
magnetic field; $\mu_{0}$ is the magnetic permeability of free space, 
$p$ is the perturbation of kinetic plasma pressure and ${\bf b}$ is the perturbation of the 
magnetic field. Let us introduce the Eulerian perturbation of the total pressure: 
${\rm P_{\rm T}}:=p+{\bf B_{0}}{\bf b}/\mu_{0}$. Because the equilibrium variables depend on 
radius only, for normal-mode analysis in cylindrical geometry we Fourier decompose 
the perturbed variables with respect to $\varphi$ and $z$ through,
\begin{eqnarray*}
\xi, {\rm P_{\rm T}} \sim \mbox{exp}[i(kz+m\varphi-\omega{t})] \ . 
\end{eqnarray*}
Here $\omega$ is the mode frequency, $m$ is the azimuthal order of the mode, 
$k$ is the longitudinal (axial) wavenumber. Then, after some algebra using the 
ideal linear MHD equations, one can find that $\xi_{\rm r}$ and 
${\rm P_{\rm T}}$ satisfy a system of two first-order ordinary differential 
equations \citep[see e.g.][]{app74, sak91},
\begin{eqnarray}
D\frac{\rm d}{{\rm d}r}(r\xi_{\rm r}) & = & C_{1}r\xi_{\rm r}-rC_{2}{\rm P_{\rm T}} \ , \mathrm{and}
\label{re_eq.4} \\
D\frac{\rm d}{{\rm d}r}{\rm P_{\rm T}} & = & \frac{1}{r}C_{3}r\xi_{\rm r}-C_{1}{\rm P_{\rm T}} \ . 
\label{re_eq.5}
\end{eqnarray}

The choice of variables above does not give much insight into the governing equation of 
Alfv\'en waves. We may need to work on further for a more suitable choice of variables, 
in order to easier identify perturbations that may be characterised as Alfv\'en waves. 
We will do this in due course. Here, the coefficient functions $D$, $C_{1}$, $C_2$ and 
$C_3$ depend on the equilibrium variables and on 
the frequency $\omega$. In general, these coefficient functions 
are given as follows:
\begin{eqnarray}
D & = & \rho_{0}(\omega^2-\omega_{\rm A}^2)C_{\rm 4}, \label{re_eq.6} \\
C_{\rm 1} & = & \frac{2B_{0\varphi}}{\mu_{0}r}\left(\omega^4B_{0\varphi}-\frac{m}{r}f_{\rm B}C_{\rm 4}\right), \label{re_eq.7} \\
C_{\rm 2} & = & \omega^4-\left(k^2+\frac{m^2}{r^2}\right)C_{\rm 4}, \label{re_eq.8} \\ 
C_{\rm 3} & = & \rho_{0}D\left(\omega^2-\omega_{\rm A}^2+\frac{2B_{0\varphi}}{\mu_{0}\rho_{0}}
\frac{\rm d}{{\rm d}r}\left(\frac{B_{0\varphi}}{r}\right)\right)+4\omega^4\left
(\frac{B_{0\varphi}^2}{\mu_{0}r}\right)^2-  \nonumber \\ 
& & \rho_{0}C_{\rm 4}\frac{4B_{0\varphi}^2}{\mu_{0}r^2}\omega_{\rm A}^2, \label{re_eq.62} 
\end{eqnarray}
where
\begin{eqnarray}
C_{\rm 4}=\left(C_{\rm S}^2+V_{\rm A}^2\right)\left(\omega^2-\omega_{\rm C}^2\right), \label{re_eq.61} 
\end{eqnarray}
and
\begin{eqnarray*}
&&C_{\rm S}^2=\gamma\frac{p_{0}}{\rho_{0}}, \quad
V_{\rm A}^2=\frac{B_{0}^2}{\mu_{0}\rho_{0}}, \quad
B_{\rm 0}^2=B_{\rm 0\varphi}^2+B_{\rm 0z}^2, \nonumber \\
&&f_{\rm B}=\frac{m}{r}B_{0\varphi}+kB_{0z}, \quad
\omega_{\rm A}^2=\frac{f_{\rm B}^2}{\mu_{0}\rho_{0}}, \quad
\omega_{\rm C}^2=\frac{C_{\rm S}^2}{C_{\rm S}^2+V_{\rm A}^2}\omega_{\rm A}^2. \nonumber
\end{eqnarray*}
Here $C_{\rm {S}}$ is the sound speed, $V_{\rm A}$ is the Alfv\'{e}n speed, $\omega_{\rm A}$ is 
the Alfv\'{e}n frequency and $\omega_{\rm C}$ is the cusp frequency. To obtain the remaining perturbed 
variables, i.e. $\xi_{\rm \varphi}$, $\xi_{\rm z}$, $b_{\rm r}$, $b_{\rm \varphi}$, $b_{\rm z}$ and 
$p$ in 
terms of $\xi_{\rm r}$ and ${\rm P_{\rm T}}$ we will have 
\begin{eqnarray}
\rho_{\rm 0}\omega^2\xi_{\rm r} & = & \frac{\rm d}{{\rm d}r}{\rm P_{\rm T}}+\frac{2B_{\rm {0\varphi}}b_{\rm \varphi}}{\mu_{\rm
0}r}-{\rm i}\frac{f_{\rm B}b_{\rm r}}{\mu_{\rm 0}}, \\
\rho_{\rm 0}\omega^2\xi_{\rm \varphi} & = & {\rm i}\frac{m}{r}{\rm P_{\rm T}}-{\rm i}\frac{f_{\rm B}b_{\rm \varphi}}{\mu_{\rm 0}}
-\frac{1}{\mu_{\rm 0}}\frac{b_{\rm r}}{r}\frac{\rm d}{{\rm d}r}\left(rB_{\rm {0\varphi}}\right), \label{eq:re_xi_varphi}\\
\rho_{\rm 0}\omega^2\xi_{\rm z} & = & {\rm i}k{\rm P_{\rm T}}-\frac{{\rm i}f_{\rm B}b_{\rm z}}{\mu_{\rm 0}}-
\frac{b_{\rm r}}{\mu_{\rm 0}}\frac{\rm d}{{\rm d}r}B_{\rm {0z}},\\
b_{\rm r} & = & {\rm i}f_{B}\xi_{\rm r}, \label{re_eq.br}\\
b_{\rm \varphi} & = & {\rm i}k\left(B_{\rm {0z}}\xi_{\varphi}-B_{\rm {0\varphi}}\xi_z\right)-
\frac{\rm d}{{\rm d}r}\left(B_{\rm 0\varphi}\xi_{\rm r}\right), \label{eq:re_b_varphi}\\
b_{\rm z} & = & -\frac{{\rm i}m}{r}\left(B_{\rm {0z}}\xi_{\rm \varphi}-B_{\rm {0\varphi}}\xi_{\rm {z}}\right)-
\frac{1}{r}\frac{\rm d}{{\rm d}r}\left(rB_{\rm 0z}\xi_{\rm r}\right), \\
p & = & -\gamma p_{\rm 0} \left(\frac{1}{r}\frac{\rm d}{{\rm d}r}(r\xi_{\rm r})+
\frac{im}{r}\xi_{\varphi}+ik\xi_{\rm z} \right).
\end{eqnarray}
From the above equations, the dominant Alfv\'en perturbations are described by 
Equations~(\ref{eq:re_xi_varphi}) \& (\ref{eq:re_b_varphi}). Note, that the perturbations 
are coupled to the other wave modes, and one may recover the torsional 
Alfv\'en wave only if the conditions $m=0$ and ${\bf B_{0\varphi}}=0$ (i.e. no 
equilibrium twist) are concurrently satisfied. Straightforward algebra yields the 
following expressions for the perturbed quantities as functions 
of the radial component of the displacement vector ($\xi_{\rm r}$) and the 
Eulerian perturbation of total pressure (${\rm P_{\rm T}}$):
\begin{eqnarray}
\xi_{\rm \varphi} & = &
\frac{\rm i}{D}\left(\frac{m}{r}C_{\rm 4}-\frac{f_{\rm B}B_{\rm {0\varphi}}}{\mu_{0}\rho_{\rm 0}}
\omega^2 \right){\rm P_{\rm T}}+ \\
& & \frac{{\rm i}f_{\rm B}B_{\rm {0\varphi}}}{\mu_{\rm 0}\rho_{\rm 0}\omega^2D}\left(C_{\rm 1}- \frac{2}{r}
C_{\rm 4}\left(\rho_{\rm 0}\omega^2-\frac{m}{r}\frac{f_{\rm B}}{\mu_{\rm 0}}B_{\rm
{0\varphi}}\right)\right)\xi_{\rm r},\\ \nonumber
\xi_{\rm z} & = & 
\frac{\rm i}{D}\left(kC_{\rm 4}-\frac{f_{\rm B}B_{\rm {0z}}}{\mu_{\rm 0}\rho_{\rm 0}}
\omega^2\right){\rm P_{\rm T}}+\\
& & \frac{{\rm i}f_{\rm B}B_{\rm {0z}}}{\mu_{\rm 0}\rho_{\rm 0}\omega^2D}\left(C_{\rm 1}+
\frac{2}{r}C_{\rm 4}\frac{m}{r}
\frac{f_{\rm B}}{\mu_{\rm 0}}B_{\rm {0\varphi}}\right)\xi_{\rm r}, \\
b_{\rm \varphi} & = & -\frac{k}{\rho_{\rm 0}(\omega^2-\omega_{\rm A}^2)}\left(g_{\rm B}{\rm P_{\rm T}}-
\frac{2f_{\rm B}B_{0\varphi}B_{0\rm z}}{\mu_{\rm 0} r}\xi_{\rm r}\right)-\frac{\rm d}{{\rm d}r}
\left(B_{\rm 0\varphi}\xi_{\rm r}\right), \\
b_{\rm z} & = & \frac{m}{r\rho_{\rm 0}(\omega^2-\omega_{\rm A}^2)}\left(g_{\rm B}{\rm P_{\rm T}}-
\frac{2f_{\rm B}B_{0\varphi}B_{0\rm z}}{\mu_{\rm 0} r}\xi_{\rm r}\right)-
\frac{1}{r}\frac{\rm d}{{\rm d}r}\left(rB_{\rm 0z}\xi_{\rm r}\right), \\
\end{eqnarray}
where
\begin{eqnarray}
&&g_{\rm B}=\frac{mB_{\rm 0z}}{r}-kB_{\rm {0\varphi}}. \\ \nonumber
\end{eqnarray}
Note, that the radial perturbed component of the magnetic field $b_{\rm r}$ 
is determined directly by Equation (\ref{re_eq.br}). In many solar applications 
it is more practical to use perturbations parallel $(\parallel)$ and perpendicular 
$(\perp)$ to magnetic field lines {\it within} constant magnetic surfaces. 
These perturbations will represent longitudinal and Alfv\'en perturbations. For 
example, in the case of no magnetic twist (i.e. $A=0$) and for axially symmetric 
perturbations only (i.e. $m=0$) the perpendicular perturbations of the magnetic 
field and velocity within the constant magnetic surfaces would 
describe the torsional Alfv\'em waves. In the geometry of the present paper 
such surfaces are concentric cylinders centred about the $z$-axis. We 
introduce the components of the displacement vector and the perturbed 
magnetic field parallel ($\xi_{\parallel}$, $b_{\parallel}$) and 
perpendicular ($\xi_{\perp}$, $b_{\perp}$) to the background equilibrium 
magnetic field lines within the constant magnetic surfaces given by,  
\begin{eqnarray}
\left(\xi_{\parallel},b_{\parallel}\right) & = & \left(\left(\xi_{\rm \varphi},b_{\rm \varphi}\right)B_{\rm 0\varphi}+
\left(\xi_{\rm z},b_{\rm z}\right)B_{\rm 0z}\right)/B_{\rm 0}, \\
\left(\xi_{\perp},b_{\perp}\right) & = & \left(\left(\xi_{\rm \varphi},b_{\rm \varphi}\right)B_{\rm 0z}-
\left(\xi_{\rm z},b_{\rm z}\right)B_{\rm 0\varphi}\right)/B_{\rm 0}.
\end{eqnarray}
These components are related to $\xi_{\rm r}$ and ${\rm P_{\rm T}}$ as, 
\begin{eqnarray}
\rho_{\rm 0}(\omega^2-\omega_{\rm C}^2)\xi_{\|} & = & \frac{{\rm i}f_{\rm B}}{B_{0}}\frac{C_{\rm S}^2}
{C_{\rm S}^2+V_{\rm A}^2}\left({\rm P_{\rm
T}}-\frac{2B_{0\varphi}^2}{\mu_{\rm 0} r}\xi_{\rm r}\right), \\ \nonumber
\rho_{\rm 0}(\omega^2-\omega_{\rm A}^2)b_{\|} & = & \frac{g_{\rm B}}{B_{\rm 0}}
\left(g_{\rm B}{\rm P_{\rm T}}-\frac{2f_{\rm B}B_{\rm {0\varphi}}B_{\rm {0z}}}{\mu_{\rm 0}r}\xi_{\rm r}\right)- \\ \nonumber
& & \frac{\rho_{\rm 0}(\omega^2-\omega_{\rm A}^2)}{B_{\rm 0}}\left(B_{\rm 0}^2
\frac{\rm {d}}{{\rm d}r}\xi_{\rm r}+\frac{B_{\rm 0z}^2\xi_{\rm r}}{r}+\xi_{\rm r}\frac{\rm d}{{\rm d}r}
\frac{B_{\rm 0}^2}{2}\right), \\ \nonumber
\rho_{\rm 0}(\omega^2-\omega_{\rm A}^2)\xi_{\bot} & = & \frac{\rm i}{B_{\rm 0}}\left(g_{\rm B}{\rm P_{\rm T}}-
\frac{2f_{\rm B}B_{0\varphi}B_{0\rm
z}}{\mu_{\rm 0} r}\xi_{\rm r}\right), \\ \nonumber
\rho_{\rm 0}(\omega^2-\omega_{\rm A}^2)b_{\bot} & = & -\frac{f_{\rm B}}{B_{\rm 0}}
\left(g_{\rm B}{\rm P_{\rm T}}-\frac{2f_{\rm B}B_{\rm {0\varphi}}B_{\rm {0z}}}{\mu_{\rm 0}r}\xi_{\rm r}\right)+ \\ \nonumber
& & \frac{\rho_{\rm 0}(\omega^2-\omega_{\rm A}^2)}{B_{\rm 0}}\left(
\frac{B_{\rm {0\varphi}}B_{\rm {0z}}}{r}+B_{\rm {0\varphi}}\frac{\rm d}{{\rm d}r}B_{\rm {0z}}
-B_{\rm {0z}}\frac{\rm d}{{\rm d}r}B_{\rm {0\varphi}}\right)\xi_{\rm r}.
\end{eqnarray}

\begin{figure*}[h!]
\centerline{
\mbox{\includegraphics[scale=0.35]{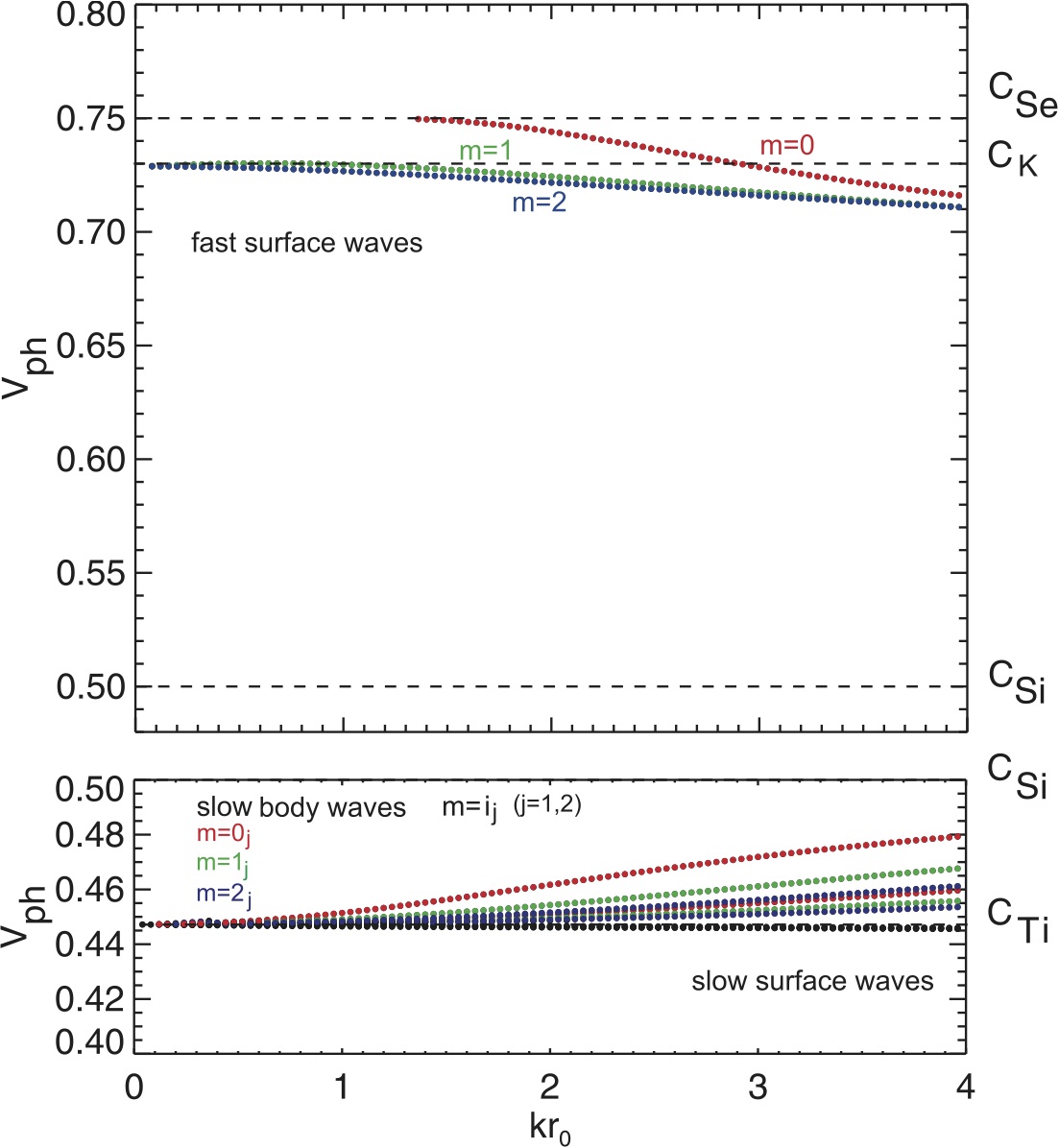}}
\mbox{\includegraphics[scale=0.35]{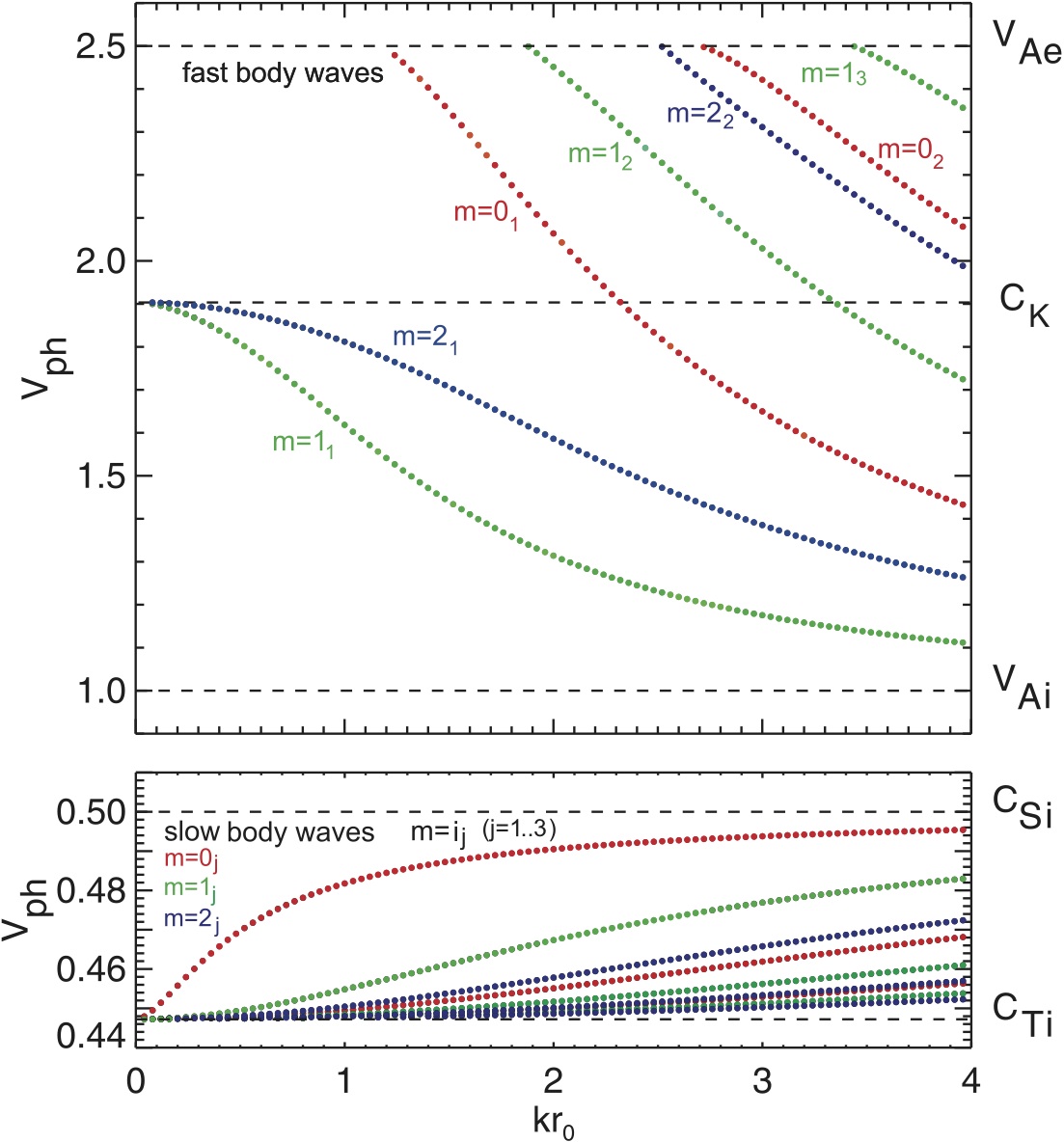}}
 }
\caption{The dimensionless phase speed $(V_{\rm {ph}})$ as function of the dimensionless 
 wavenumber $(kr_{\rm 0})$ under (a) typical photospheric 
 (i.e $V_{\rm {Ai}}>C_{\rm {Se}}>C_{\rm {Si}}>V_{\rm {Ae}}$) and 
 (b) coronal conditions (i.e $V_{\rm {Ae}}>V_{\rm {Ai}}>C_{\rm {Si}}>C_{\rm {Se}}$) for an untwisted
 magnetic flux tube are shown at the (a) left and (b) right panels, respectively. 
 Here (a) $C_{\rm {Se}}=0.75V_{\rm {Ai}}$, 
 $V_{\rm {Ae}}=0.25V_{\rm {Ai}}$, $C_{\rm {Si}}=0.5V_{\rm {Ai}}$ for the photospheric case and 
 (b) $C_{\rm Se}=0.25V_{\rm {Ai}}$, $V_{\rm{Ae}}=2.5V_{\rm {Ai}}$, $C_{\rm {Si}}=0.5V_{\rm {Ai}}$ 
 for the solar corona. In the left panel three modes of the fast surface waves that are bounded within 
 $[C_{\rm {Si}},C_{\rm {Se}}]$ and the infinite number of the slow body waves bounded within 
 $[C_{\rm {Ti}},C_{\rm {Si}}]$ are  plotted. 
 The slow kink, sausage, etc. surface waves are very close to each other (just under $C_{\rm {Ti}}$). 
 In the right panel sausage (m=0), kink (m=1) 
 and fluting (m=2) modes are shown. 
 The infinite number of the fast body waves are bounded within $[V_{\rm {Ai}},V_{\rm {Ae}}]$ and 
 the infinite number of the slow body waves are bounded within $[C_{\rm {Ti}},C_{\rm {Si}}]$.
 $m=i_{j}$, where the $i$ refers to 
 the mode (sausage, kink, etc.) and $j$ refers to the $j$-th branch of the zeroes of eigenfunctions in the radial
 direction. Only three branches of each mode of the infinitely many slow and fast body waves are plotted. }
\label{fig:re_dr-solution}
\end{figure*}

These governing equations, in various alternative formats, were derived previously in 
various contexts by e.g. \cite{hai58}, \cite{goe71}, \cite{app74}, \cite{goo91}, \cite{sak91} 
and \cite{erd-fed06, erd-fed07, erd-fed10}.

\subsection{Derivation of the General Dispersion Equation}
\label{subsec:dr}
Applying the boundary conditions to the inside and outside solutions of the 
governing equations yields the required general dispersion relation that will 
determine the existing wave modes of the system. The continuity condition for 
the radial component of the displacement,
\begin{eqnarray*}
&& \left. \xi_{\rm {ir}}\right|_{r=r_{\rm 0}}=\left. \xi_{\rm {er}} \right|_{r=r_{\rm 0}},  
\end{eqnarray*}
and the continuity of pressure,
\begin{eqnarray*}
&& \left. {\rm P_{\rm {Ti}}}-\frac{B_{\rm {0i\varphi}}^2}{\mu_{0}r_{\rm 0}}\xi_{{\rm i}r} \right|_{r=r_{\rm 0}}=
\left. {\rm P_{\rm Te}}|_{r=r_{\rm 0}} \right. \\
\end{eqnarray*}
provide the desired dispersion relation:
\begin{eqnarray}
D_{\rm e}\frac{r_{0}}{m_{\rm {0e}}}\frac{K_m(m_{\rm {0e}}r_0)}{K_m^{'}(m_{\rm {0e}}r_0)}=-\frac{A^2r_0^2}{\mu_0^2}+
D_{\rm i} r_0^2\frac{( 1-\alpha^2)}{m\left ( 1-\alpha\right)+2\mathrm{x_0}
\frac{\mathit{M}^{'}\left ( a,b,\mathrm{x_0}\right )}{\mathit{M}\left ( a,b,\mathrm{x_0}\right )}}. \label{re_eq.22}
\end{eqnarray}
The dash denotes the derivative of the Kummer's function evaluated at $\mathrm{x}=\mathrm{x}_0$, where
\begin{equation}
\mathrm{x_{\rm 0}}=\frac{n}{k}E^{1/2}k_{\alpha}^2r_{\rm 0}^2.
\end{equation}

Equation (\ref{re_eq.22}) is the {\it general dispersion equation} for linear MHD waves in a 
magnetically uniformly twisted compressible flux tube embedded in a compressible and 
uniform straight magnetic environment. To the best of our knowledge, 
it is not possible to solve analytically Equation (\ref{re_eq.22}) in general terms. 
However, at least certain checks can be  carried out, e.g. applying the 
incompressible limit should recover the dispersion relation of \cite{ben99} or
 \cite{erd-fed06}. Another limiting case 
is $B_{0\varphi}\rightarrow 0$ where the well known dispersion relation Equations (8a, b) of 
\cite{edw_rob83} must be recovered. These particular cases of analytical 
checks reassure us that the derivations are consistent.  Although we would 
prefer further analytical insight, such approach seems to be limited 
by the lack of existing suitable asymptotic expansions for Kummer functions 
$\mathit{M}\left ( a,b,\mathrm{x_0}\right )$ and $\mathit{M}^{'}\left ( a,b,\mathrm{x_0}\right )$. 
We shall here only recall a typical 
graphical solution of the dispersion relation, revealing all the linear MHD modes. 

In Figure~\ref{fig:re_dr-solution} the results of calculations of the dispersion curves derived from 
Equation (\ref{re_eq.22}) are presented. One can observe here the very rich structure of linear 
MHD modes. Note, the Alfv\'en waves are just a single line,
as all the modes have the same phase speed. The modes can often only be 
distinguished from other magneto-acoustic (e.g. kink) modes, if one carefully determines the 
polarisation. This can be rather hard, given the currently available observational tools. 

\section{Concluding remarks}
There are at least three fundamental questions that must be resolved in order to 
understand the heating of the solar (and stellar) atmosphere: Where is the energy 
generated? How does the generated energy propagate from the energy reservoir 
to the solar atmosphere, spanning from the warm chromosphere to the corona? 
How does the transported energy efficiently dissipate in the solar atmosphere in 
order to maintain its multimillion-Kelvin temperature?
There are numerous examples of slow and fast MHD waves observed in 
the solar atmosphere. However, their energy density does not seem sufficient  
for coronal heating or powering the solar wind. On the other hand, among 
MHD wave theorists, Alfv\'en waves still remain the ``Holy Grail'' of coronal 
heating, solar wind acceleration and solar magneto-seismology \citep{Ban_etal07}. 

The applicability of MHD wave research to magneto-seismology is a relatively recent 
development. The main idea here is to determine information about the MHD 
waveguide in which the waves propagate. This is analogous to the techniques of 
seismology, a popular method of investigating the sub-surface structure of the Earth. 
The difference here is that in a magnetised plasma there is a 
plentiful abundance of waves in existence, offering us various aspects of 
seismology. The frequency (or period) measurements of MHD waves are often 
used to determine density and magnetic stratification, both in the longitudinal and  
transversal directions of the waveguide. More recently, however, studies of 
the spatial distribution of waves and their intrinsic characteristics have been suggested, 
offering an independent method to determine diagnostic information about the 
waveguide \citep{erd_ver07, ver_etal07}. Magneto-seismology has a 
novel application to the reconstruction of MHD waveguides. This is perhaps an even 
more important aspect compared to the others. 
The idea here is that by studying the properties of MHD waves and oscillations, 
their distribution in the waveguide (assuming here an adequate spatial and 
temporal resolution) may help construct detailed magnetic surface topology. 
The idea has been outlined in a recent paper by 
\cite{ver_etal10} and \cite{fed_etal11b}.

There have been some suggestions that transversal waves may be considered 
Alfv\'enic in nature providing that they are incompressible, exhibit no intensity 
fluctuations and display periodic displacements with the magnetic tension as the 
restoring force \citep{RefAu}. The terminology of Alfv\'enic is, however, not entirely 
clear as such waves do not normally exist as an eigenvalue of the linear 
MHD equations in structured plasmas: on the other hand many magnetic oscillating structures in the corona and below can not be described by pure linear MHD eigenmodes. Given the ambiguity, it may be better 
to refer to these waves as transversal \citep{RefAv}. Further, refined observations, with the 
use of e.g. spectro-polarimetric measurements, may be able to clarify the wave mode 
identity with greater accuracy. 
 
According to \cite{RefAt} and \cite{RefT}, signatures of Alfv\'en waves were 
observed with the use of the XRT and SOT instruments, respectively. Because 
these jets are increases in local density, if their length is short relative to the 
wavelength of Alfv\'en perturbations, they can slide along magnetic field lines, 
and act as test particles (see the yellow-red blob in Fig.~\ref{fig:re_aw}b) 
in the plasma that has almost no resistivity, just like bobsleds would run and 
slide along a prebuilt track. 
However, these observations also raise concerns about the applicability of the 
classical concept of a magnetic flux tube in the very dynamic solar atmosphere, 
where these sliding jets were captured. In a classical magnetic flux tube, 
propagating Alfv\'en waves along the tube would cause torsional oscillations 
(Fig.~\ref{fig:re_aw}a). In this scenario, the only observational signature of 
Alfv\'en waves would be spectral line broadening, as observed by \cite{RefZ}. 
 On the other hand, if these classical flux tubes did indeed exist, 
then the observations of \cite{RefT} would be interpreted as kink waves 
(i.e. waves that displace the axis of symmetry of the flux tube like an S-shape). 
More detailed observations are needed, so that a full three-dimensional picture of 
wave propagation could emerge. 

Numerical simulations of a two-dimensional stratified VAL-atmosphere model 
\citep{RefAe}, embedded in a horizontal magnetic field to mimic solar prominences and 
driven by global photospheric motions, will also result into observational signatures very 
similar to those reported by \cite{RefAr}, and in Figure~2 and the online movie S1 of 
\cite{erd-fed07}. In particular, note the similarity between the time-distance plot (Fig.~2a) 
of the illustrative forward modelling simulations in \cite{erd-fed07} and Figure~3 of 
\cite{RefAr}. However, in the case of this numerical forward modelling experiment 
the oscillations found are clearly not Alfv\'enic, 
but are rather transversal kink oscillations. 

It is worth mentioning what may be some key 
directions where MHD, in particular, Alfv\'en wave research may develop. 
Among the many possibilities is the development of Alfv\'en (and in parallel other 
MHD) wave theory in dynamic plasma. Observations indicate that waveguides 
change, often considerably, during the time that the oscillations reported are 
guided by them \citep{asc08}. As long as the characteristic time scale of the 
changes in the background plasma are small or comparable to the time scales 
of waves crossing the structure, some initial theoretical 
progress has already been made \citep{mor09,mor10,mor_etal10,mor_etal11,rud11a,rud11b}. 
It is a real challenge to develop wave theory in dynamic plasmas. Another fascinating 
aspect of developments will be the advancement of (non)linear  Alfv\'en and 
magneto-acoustic waves where the geometry and structuring of the waveguide is more 
complex but more realistic. Significant technological improvements associated in High 
Performance Computing will allow detailed MHD wave studies of the coupled solar 
interior-atmosphere system as a whole. Such studies will shed ample light into the 
true nature of Alfv\'en and other MHD wave dynamics and heating 
\citep{bog03,car06,erd_etal07,fed_etal09,kho_etal08,fed_etal11a}. 

Last, but by no means least, we will provide some observational goals that will 
hopefully assist the detection, disambiguation, and study of 
solar oscillations. High-cadence imagers, such as the ROSA \citep{Jess10} 
instrument, will need to be employed with high spatial, temporal, {\it{and}} spectral 
resolution facilities, such as the IBIS \citep{Cav06} and CRISP \citep{Sch08} 
two-dimensional spectro-polarimeters. Simultaneous observations between the ground-based 
instruments and state-of-the-art space-borne facilities such as SDO,  and the planned IRIS 
(Interface Region Imaging Spectrograph) can provide the necessary intensity, magnetic-field, 
and Doppler signals required to unveil the specific wave modes as they travel through the 
solar atmosphere. The Helioseismic and Magnetic Imager (HMI) on SDO provides 
dopplegrams, continuum filtergrams, line-of-sight and vector magnetograms at the solar 
surface, thus allowing the sites of photospheric oscillations to be identified (albeit at a lower 
spatial resolution). The Atmospheric Imaging Assembly (AIA) provides the coronal 
magnetic field configuration, which can be compared to that obtained by HMI. IRIS in particular 
will obtain high spatial (250~km) and temporal resolution (1~sec) spectra and images in 
ultraviolet wavelengths, focusing on the upper chromosphere and transition region. The 
combination of simultaneous imaging and spectroscopy provided by IRIS will allow us to 
disentangle the velocity and intensity signal of torsional, kink and sausage wave modes 
as they travel through the complex interface between the photosphere and the corona. 
High resolution limb spectro-polarimetry of chromospheric structures such as spicules 
and prominences remains a major challenge. The low photon statistics at the solar limb, 
limit the use of Adaptive Optics and the number of photons required for the measurement 
of accurate Stokes profiles. These difficulties will be overcome with the upcoming 
Advanced Technology Solar Telescope \citep{Kei10}, with a 4~m primary mirror, will allow 
structures in the Sun's atmosphere to be examined on unprecedented spatial scales of $\approx$20~km.

\begin{acknowledgements}

DBJ thanks the Science and Technology Facilities Council (STFC) for a Post-Doctoral Fellowship. 
RE acknowledges M. K\'eray for patient encouragement and is grateful to NSF Hungary (OTKA K83133). RE also thanks for the hospitality received at the Dept. of Astronomy, E\H{o}tv\H{o}s University, Budapest, Hungary, where part of his contribution to the review paper was developed. We would like to thank the anonymous referees for their comments and suggestions on the manuscript. 
\end{acknowledgements}



\end{document}